\documentclass[lettersize,journal]{IEEEtran}

\usepackage{enumerate}
\usepackage{algorithm}
\usepackage{float}
\usepackage{color}
\usepackage{makeidx}
\usepackage{bm}
\usepackage{bbm}
\usepackage{url}
\usepackage{steinmetz}
\usepackage[algo2e,linesnumbered,lined,boxed,commentsnumbered]{algorithm2e}
\usepackage{varwidth}
\usepackage{siunitx}
\usepackage{balance}
\usepackage{enumerate}

\usepackage{amsmath,amsthm}
\usepackage[noend]{algpseudocode}
\usepackage{lipsum}
\usepackage{soul}
\usepackage{dsfont}

\usepackage{amsfonts}
\usepackage{times}
\usepackage{graphicx}
\usepackage{latexsym}
\usepackage{dsfont}
\usepackage{amssymb}
\usepackage{cite}
\usepackage{verbatim}
\usepackage{subfigure}

\newtheorem{theorem}{Theorem}

\newtheorem{remark}[theorem]{Remark}

\newcommand{\figref}[1]{{Fig.}~\ref{#1}}

\def\bb0{{\mathbb{0}}}

\def\ba{{\mathbf{a}}}
\def\bb{{\mathbf{b}}}

\def\bff{{\mathbf{f}}}

\def\bh{{\mathbf{h}}}

\def\bm{{\mathbf{m}}}
\def\bn{{\mathbf{n}}}

\def\bt{{\mathbf{t}}}

\def\by{{\mathbf{y}}}
\def\bz{{\mathbf{z}}}
\def\b0{{\mathbf{0}}}

\def\bI{{\mathbf{I}}}

\def\bR{{\mathbf{R}}}

\def\bbE{{\mathbb{E}}}

\def\sf0{{\mathsf{0}}}

\usepackage{epstopdf}

\newcommand{\sref}[1]{{Section}~\ref{#1}}

\DeclareMathOperator*{\argmax}{arg\,max}

\newcommand{\abs}[1]{\lvert#1\rvert}
\newcommand{\absq}[1]{\left\lvert#1\right\rvert^2}

\def \bpsi{\boldsymbol{\psi}}

\begin{document}

\title{Enabling Cell-Free Massive MIMO Systems with Wireless Millimeter Wave Fronthaul}

\author{Umut Demirhan and Ahmed Alkhateeb \thanks{The authors are with the School of Electrical, Computer and Energy Engineering, Arizona State University, Tempe, AZ, 85287 USA. (Email: udemirhan, alkhateeb@asu.edu).}}

\maketitle

\begin{abstract}
	Cell-free massive MIMO systems have promising data rate and coverage gains. These systems, however, typically rely on fiber based fronthaul for the communication between the central processing unit and the distributed access points (APs), which increases the infrastructure cost and installation complexity. To address these challenges, this paper proposes two architectures for cell-free massive MIMO systems based on wireless fronthaul that is operating at a higher-band compared to the access links. These dual-band architectures ensure high data rate fronthaul while reducing the infrastructure cost and enhancing the deployment flexibility and adaptability. To investigate the achievable data rates with the proposed architectures, we formulate the end-to-end data rate optimization problem accounting for the various practical aspects of the fronthaul and access links. Then, we develop a low-complexity yet efficient joint beamforming and resource allocation solution for the proposed architectures based on user-centric AP grouping. With this solution, we show that the proposed architectures can achieve comparable data rates to those obtained with optical fiber-based fronthaul under realistic assumptions on the fronthaul bandwidth, hardware constraints, and deployment scenarios. This highlights a promising path for realizing the cell-free massive MIMO gains in practice while reducing the infrastructure and deployment overhead.
\end{abstract}

\section{Introduction}
\textbf{Can we realize the cell-free massive MIMO gains with wireless fronthaul? } This paper attempts to answer this important question. Cell-free massive MIMO systems are promising uniform coverage and high data rate gains, even for scenarios with very dense users \cite{ngo2017,interdonato2019ubiquitous}. A critical challenge, however, with the current cell-free massive MIMO systems is its reliance on optical fiber-based fronthaul for distributing data and synchronization signals from the central processing unit (CPU) to the access points (APs). This highly increases its infrastructure cost and installation time, and limits the flexibility and  scalability of the cell-free massive MIMO deployment.  To address these challenges, the paper proposes an alternative cell-free massive MIMO architecture with higher-band fronthaul, e.g., a millimeter wave (mmWave) or terahertz (THz) fronthaul for a sub-6GHz cell-free massive MIMO system. The use of higher frequency band fronthaul has two key advantages: (i) The large bandwidth available at the higher frequency bands provides a high data rate fronthaul and (ii) the higher-band fronthaul signals (lower wavelength) have high capability to synchronize the sub-6GHz AP transceivers. For this proposed architecture, the paper investigates whether it can achieve comparable achieve rates to those obtained by classical fiber-fronthaul based  cell-free massive MIMO systems. 

\subsection{Related Works}

Motivated by its potential to increase the data rate and manage the multi-user interference, the idea of having distributed antennas simultaneously serving the same users have been previously investigated in the network multiple-input multiple-output (MIMO) \cite{venkatesan2007network}, distributed MIMO \cite{wang2013spectral}, and coordinated multipoint with joint transmission (CoMP-JT) \cite{irmer2011coordinated, marsch2011coordinated}. These earlier approaches did not scale well with the increasing number of users and large number of access points (APs), due to the channel estimation and feedback overhead. To overcome this limitation, in \cite{ngo2017, ngo2017total, nayebi2017precoding,Buzzi2017,Alrabeiah2019c}, the authors investigated cell-free massive MIMO, where reciprocity of the time-division duplexing (TDD) is leveraged to estimate the downlink channels directly from the joint uplink pilot transmissions. In the proposed cell-free optimization framework, the imperfection of channel state information (CSI) was taken into account, and only the long-term channel coefficients were utilized for the power allocation. Despite the interesting data rate and coverage gains, the realization of cell-free massive MIMO networks suffers from the limitations and high infrastructure cost of the CPU-AP fiber links. To reduce cost, cheaper wire-link alternatives, with limited capacity, were considered. For example, \cite{bashar2019max} investigated cell-free performance with quantized fronthaul transmissions. However, the approaches in \cite{bashar2019max, bashar2018, femenias2019cell, femenias2019reduced} did not eliminate the wired connection requirement, which is still associated with high infrastructure cost, high installation time, and limited deployment flexibility. To clarify, the joint consideration of the wireless fronthaul and cell-free massive MIMO networks has not been studied in the literature.

Another relevant line of work is presented in the different network architectures that adopted wireless fronthaul/backhaul such as small-cells \cite{zhu2016, hao2017small}, ultra dense networks \cite{gao2015mmwave}, heterogeneous networks \cite{bogale2016massive} and the cloud radio access networks (C-RAN) \cite{stephen2017joint, hu2017joint, hu2018joint}. In \cite{zhu2016, hao2017small, gao2015mmwave, bogale2016massive}, wireless fronthaul was adopted to show the potential advantages of the mmWave backhaul. This was however, limited to scenarios with no cooperation between basestations. In C-RAN architectures, multiple (normally a few) base stations coordinate to serve the users. For example, in \cite{stephen2017joint}, the time-frequency resources of an OFDMA based C-RAN system consisted of a few base stations were optimized to maximize the weighted sum-rate of large number of users. The solution in \cite{stephen2017joint} assumed the fronthaul transmission to a single base station at a time. To overcome this limitation, the work in \cite{hu2017joint} proposed a multicast beamforming for the downlink ultra dense C-RAN, and designed user-centric clusters of the base stations for the transmissions. The work in \cite{stephen2017joint, hu2017joint}, however, was limited to architectures with a few number of base stations. This is partially due to the large channel estimation/feedback overhead associated with the solutions in \cite{stephen2017joint, hu2017joint}, which limit them from scaling to large number of antennas. This motivates the research for new approaches to enable the potential gains of cell-free massive MIMO systems with the adoption of the wireless fronthaul.

\subsection{Contributions}
In this work, we propose a new architecture for cell-free massive MIMO systems where the fronthaul is implemented using high-frequency (e.g., mmWave) wireless links  to serve access links at lower frequency (e.g., at a sub-6GHz band). The proposed architecture has two key motivations: (i) The larger available bandwidth at the high frequency band ensures high data rates for the fronthaul and (ii) the small wavelength of the high-frequency signals provides high synchronization accuracy for the access points that are operating at a lower frequency band. Further, the wireless fronthaul enables a modular, flexible and scalable architecture with low infrastructure cost and low installation time. With all these potential gains, an important question is whether this architecture is capable of achieving comparable data rate gains to those achieved with optical fiber-based cell-free massive MIMO systems?  To answer this question, we develop an efficient communication scheme, analyze its performance, and draw important insights about the proposed architecture and data rate optimization approaches. The main contributions of the paper can be summarized as follows: 
\begin{itemize}
	\item \textbf{Proposing an efficient architecture for cell-free massive MIMO systems based on higher-frequency wireless fronthaul}. The proposed architecture has the potential of ensuring higher data rate fronthaul and high synchronization accuracy for the distributed APs. Further, it requires low infrastructure cost and low installation time and provides interesting flexibility and adaptability gains for cell-free massive MIMO systems.

	\item \textbf{Developing an efficient communication model for the wireless fronthaul based cell free massive MIMO system}. This model assigns a group of APs to each user and optimizes the multicast beamforming at the central processing unit to simultaneously serve the AP group of each user. The adopted system model accounts for the practical constraints on the higher-frequency band (mmWave) beamforming architectures \cite{Alkhateeb2014c}. Specifically, the CPU applies analog beamforming. Given the spatial multiplexing constraints of the analog beamforming, we use time-division multiple-access (TDMA) to multicast the message of each user-centric group.

   \item \textbf{Formulating the end-to-end data rate maximization problem for the proposed architecture and communication model}. The optimization problem takes the wireless fronthaul, access channels, and AP grouping into account. In particular, it aims to determine the user-centric AP group selection, fronthaul beamforming vectors, TDMA schedule, and AP power coefficients to maximize the end-to-end data rates.

	\item \textbf{Developing near-optimal end-to-end data rate maximization solution for the proposed architecture}. 
	The solution adopts an iterative group selection algorithm, which is coupled with the fronthaul and access channel data rate maximization sub-problems. Specifically, in each iteration, the group size and AP selection are determined based on the channel estimates and then the fronthaul/access rates are optimized for the given groups. 
	\item \textbf{Extending the architecture and developed solutions to the mixed wireless/wired fronthaul case.} A mixed-fronthaul architecture with  wire-connected AP clusters (for example through a radio stripe \cite{interdonato2019ubiquitous}) is proposed. In this architecture, only the leader AP in each wire-connected cluster has a wireless fronthaul link with the CPU. The proposed data rate maximization solutions are generalized to account for the mixed-fronthaul. 
\end{itemize}
The proposed solutions are extensively evaluated using numerical simulations which draw important insights into the performance of the proposed cell-free architectures. Based on the results, the high-frequency wireless fronthaul can provide sufficient data-rates for the cell-free massive MIMO by taking advantage of the larger bandwidth availability. With the fully wireless architecture, it is possible to achieve data rates with only $10-20\%$ degradation compared to the fiber fronthaul based cell-free massive MIMO architectures. Further, the mixed-fronthaul architecture significantly reduces the bandwidth requirements and improves the data-rates. In our simulations, the mixed-fronthaul architecture enabled data rates very similar to the fiber fronthaul based solutions with reasonable fronthaul bandwidth requirements.

\textbf{Notation}: We use the following notation throughout this paper: $\ba$ is a vector, $a$ is a scalar, $\mathcal{A}$ is a set of scalars, and $\boldsymbol{\mathcal{A}}$ is a set of vectors or sets. $\abs{\ba}$ is the absolute value of $\ba$, whereas $\ba^T$, $\ba^H$ are its transpose and Hermitian. $\mathcal{CN}(\bm,\bR)$ is a complex Gaussian random vector with mean $\bm$ and covariance $\bR$. $\mathcal{U}[a, b]$ is a uniform random variable in $[a, b]$. $\bbE\left[\cdot\right]$ is used to denote expectation.

\section{Wireless-Fronthauled Dual-Band \\ Cell-Free Massive MIMO: The Key Idea} \label{sec:architecture}
\begin{figure*}[!t]
	\centering
	\includegraphics[width=1.7\columnwidth]{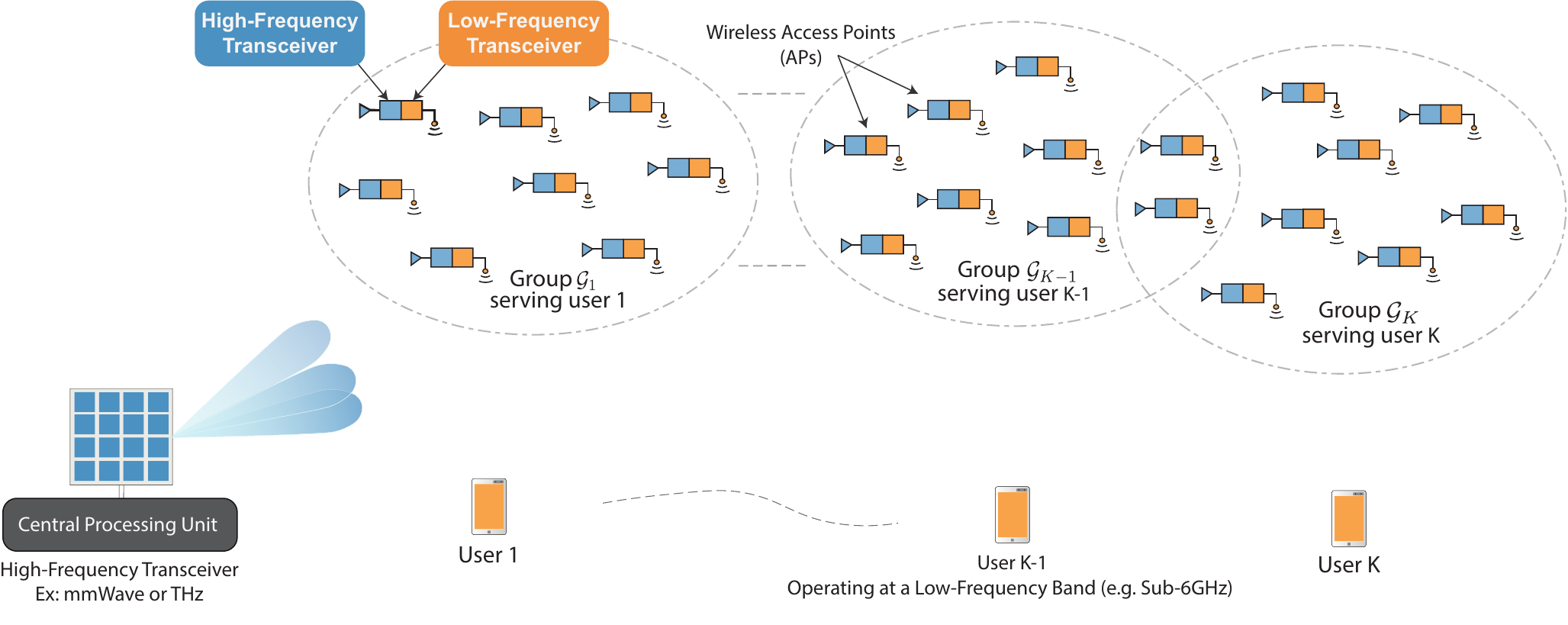}
	\caption{An illustration of the proposed architecture where a mmWave CPU provides fronthaul to the different sets of user-centric AP groups via different beamforming vectors. The wireless APs jointly serve their users through a cell-free massive MIMO access channel at sub-6GHz.}
	\label{fig:systemmodel}
\end{figure*}

Cell-free massive MIMO is a promising enabler for the high data rate and  coverage requirements in future wireless communications systems. Current cell-free massive MIMO architectures, however, assume that all the antennas are connected to the central processing unit with wired fronthaul, that is typically optical fiber \cite{pfeiffer2015next}. This complicates the deployment process of these systems and  increases its cost and installation time. To overcome these challenges, we propose an alternative architecture for cell-free massive MIMO systems that relies on a wireless fronthaul operating at a much higher frequency band compared to the access band, as depicted in \figref{fig:systemmodel}. For example, a mmWave fronthaul channel with sub-6GHz access channels or a terahertz fronthaul supporting mmWave access channel. Thanks to the wireless fronthaul, \textbf{the proposed architecture is modular, flexible, scalable, with low cost and installation time}. In particular, replacing the wired-fronthaul with a wireless connection provides modularity and flexibility in the deployment. This way, the costs and installation time can be reduced as the APs can be directly deployed, potentially without any (or with minimal) engineering assistance. Consequently, this allows easier deployment of the APs and potential deployment scalability of the system\footnote{Note that a scalability notion for cell-free massive MIMO was formally defined in\cite{bjornson2020scalable}, which aims to keep the complexity of the channel estimation, signal processing for beamforming, fronthaul signaling and power control optimization limited for asymptotically increasing number of users. It is important to note that this is different than our use of scalability, that is for the deployment of the APs.}. While the wireless fronthaul provides these advantages, it may have challenges for the data rates and synchronization, which motivates the use of a higher-frequency fronthaul. The use of a higher frequency band in the fronthaul compared to the access band has the following advantages (for the rest of the paper, and for ease of exposition, we will assume a mmWave fronthaul and a sub-6GHz access channel):

\begin{itemize}
    \item \textbf{High Data-Rate Fronthaul:} The availability of the large bandwidth at the mmWave frequency band enables the central processing unit (which is equipped with a mmWave transceiver) to support high data-rate fronthaul. This way, the crucial large bandwidth availability can be utilized for maintaining the high data rate gains of cell-free massive MIMO systems, as will be shown in \sref{sec:numresults}.
    
    \item \textbf{High Synchronization Accuracy:} Using a mmWave fronthaul (where signals have a small wavelength) to synchronize the sub-6GHz APs (where signals have a much higher wavelength) has the potential of ensuring precise clock synchronization among the APs, which is essential for the operation of the cell-free massive MIMO systems. Specifically, the synchronization of the APs can be achieved over the mmWave fronthaul with the aid of the CPU via master-slave type algorithms (e.g., \cite{shrestha2018precise}) or via network centric solutions (e.g., \cite{rogalin2014scalable}). The APs use two clock signals, one for the mmWave fronthaul and the other for the sub-6GHz access channel. The APs can synchronize their mmWave clocks and time by referring to the CPU. These signals can then be utilized for more accurate sub-6GHz synchronization by converting the mmWave (smaller wavelength) signal to a large wavelength clock\footnote{For instance, if there are two mmWave clocks that are synchronized with a $\Delta f$ frequency offset between them, the sub-6GHz clocks obtained from the these (or from the same source) can translate into a smaller frequency difference, possibly at the ratio of the carrier frequencies (e.g., for $30$ GHz and $3$ GHz carrier frequencies, it could be possible to obtain the clocks with frequency difference $\frac{\Delta f}{10}$). A similar gain is expected if a THz-based fronthaul is leveraged to synchronize a mmWave access channel. }.
\end{itemize}

Further, with regards to the mmWave fronthaul, we would like to highlight two points: (i) The cost of mmWave transceivers may currently be high. However, it is expected to decrease over time with the mass adoption and production. Further, in Section \sref{sec:connectedAPs}, we propose an extension of the architecture that requires much less number of mmWave transceivers. (ii) The potential coverage limitations of the mmWave fronthaul can be reduced with the employment of a large antenna array at the CPU, as will be considered in the following sections. In general, we envision that the proposed architecture will be promising for the scenarios where the CPU is serving APs distributed within a few hundred meters.  

Now, with the interesting gains of having wireless fronthaul based cell-free massive MIMO systems, and while mmWave fronthaul may have relatively high data rates compared to other wireless solutions, it is important to answer the question: \textbf{Can this wireless fronthaul based architecture achieve the same data rates of the fiber fronthaul based cell-free massive MIMO architectures?} In this paper, we try to answer this question. Towards this objective, we  first describe the system model of the proposed mmWave fronthaul based cell-free massive MIMO architecture in \sref{sec:systemmodel} and formulate the end-to-end (from the CPU to the users) achievable rate optimization problem in \sref{sec:problemformulation}. Then, we  develop an efficient transmission strategy for the proposed architecture in \sref{sec:solution} and discuss some important extensions in \sref{sec:connectedAPs}. Finally, the achievable rate of the proposed architecture and comparisons with the fiber-based solutions are  provided in \sref{sec:numresults}.

\section{System Model} \label{sec:systemmodel}

We consider the distributed massive MIMO system in \figref{fig:systemmodel}, where a  CPU communicates with $M$ wireless-APs over a high-frequency (e.g., mmWave) wireless fronthaul  and the wireless-APs serve $K$ user equipment (UEs) over a low-frequency (e.g., sub-6GHz) channel. We will use  $\mathcal{M}=\{1, \ldots, M\}$ and $\mathcal{K}=\{1,\ldots,K\}$ to denote the sets of  $M$ APs and $K$ UEs. For the rest of the paper, we will refer to the downlink channel from the CPU to the APs as the \textit{fronthaul channel} and the downlink channel from APs to the UEs as the \textit{access channel}. Further, for the ease of exposition and without loss of generality, we will assume that the fronthaul channel is operating over a mmWave band, with a bandwidth  $B^{\mathrm{fh}}$, and the access channel is adopting a sub-6GHz band, with a bandwidth $B^{\mathrm{ac}}$. The algorithms and results of the paper, however, can be applied to other dual-band architectures, such as terahertz fronthaul with a mmWave access channel. 

To beamform the signal to the APs,  the CPU employs an antenna array of  $N$ elements, while the APs and UEs are, for simplicity,  assumed to have single antennas.  We consider a user-centric grouping approach where the message of UE $k$ is jointly transmitted by a subset of APs, $\mathcal{G}_k \subseteq \mathcal{M}$. We denote the set of the user-centric groups as $\boldsymbol{\mathcal{G}}=\{\mathcal{G}_1, \ldots, \mathcal{G}_K\}$. Note that different groups may include the same APs since multiple APs are utilized in the transmission to every UE. For that, we will also define  $\mathcal{U}_m \subseteq \mathcal{M}$ for $m\in \mathcal{M}$ as the set of users that are being served by the AP $m$. For the distances between the CPU and $m$-th AP and between the $m$-th AP and  $k$-th user, we will use $d^{\mathrm{fh}}_m $ and $d^{\mathrm{ac}}_{mk}$ to denote them. It is important to mention here that we do not assume any knowledge about the positions or distances between the CPU, APs, or the users.

\textbf{At the Fronthaul Link:} The mmWave CPU adopts TDMA to serve the $K$ user-centric groups. In each TDMA slot, the CPU beamforms the signal towards the APs that serve one user. The duration of the TDMA slot allocated for serving APs of UE $k$ is denoted by $t_k$. Let $\bh_m \in \mathbb{C}^N$ denote the channel between the CPU and the $m$th AP. If the message intended for user $k$ is represented by $q^{\mathrm{fh}}_k \in \mathbb{C}$,  with $\bbE[\absq{q_k^{\mathrm{fh}}}] = 1$, then the received signal at the AP $m$ can be written as
\begin{equation} \label{eqn:fronthaulsignalmodel}
y^{\mathrm{fh}}_m = \sqrt{\rho^\mathrm{fh}} \bh_m^H \bff_k q^{\mathrm{fh}}_k + w^\mathrm{fh}_m,
\end{equation}
where $\rho^\mathrm{fh}$ is the normalized fronthaul transmission power and $w^\mathrm{fh}_m \sim \mathcal{CN}(0, 1)$ is the receive noise at the $m$th AP.  The vector $\bff_k \in \mathbb{C}^N$ is the CPU beamforming vector intended to focus the signal to the APs that serve user $k$. To satisfy the practical mmWave hardware constraints, we assume that the CPU adopts analog-only beamforming implemented by a network of quantized phase shifters \cite{Alkhateeb2014c}. This means that the beamforming vector $\bff_k$ can only be selected from a certain set of vectors, that we define by the codebook $\boldsymbol{\mathcal{F}}$. If each phase shifter has  $q$ bits, i.e., $2^q$ possible phase shift values defined by the set $\mathcal{Q} \in\{ 0, \frac{\pi}{2^q},  \frac{2 \pi}{2^q}, ...,   \frac{(2^q -1)\pi}{{2^q}} \}$, then we can write
\begin{equation}
\boldsymbol{\mathcal{F}} = \left\{\frac{1}{\sqrt{N}}[e^{j \phi_1}, \ldots, e^{j \phi_N}]^T : \phi_n \in \mathcal{Q}, \forall n\in \{1, \ldots, N\}\right\},
\end{equation}
and $\bff_k \in \boldsymbol{\mathcal{F}}$. We assume that the channels between the CPU and APs are available at the CPU. This is motivated by the stationarity of the CPU and APs and the channel reciprocity assumption. 

\textbf{At the Access Link:} Each AP decodes the CPU signal received by its mmWave receiver and prepares it for the transmission over the sub-6GHz access channel. Similar to \cite{ngo2017}, the APs transmit the weighted sum of the users' messages, where each message is multiplied by a different beamforming and power control coefficient. Since each AP $m$ is assumed to contribute in serving a set of users $\mathcal{U}_m$, the transmitted signal from the $m$th AP, $x_m$, can be written as
\begin{equation} \label{eqn:accesschannelsignalmodel-a}
x_{m} = \sqrt{\rho^\mathrm{ac}} \sum_{k \in \mathcal{U}_m} \sqrt{p_{mk}} f^\mathrm{ac}_{mk} q^{\mathrm{ac}}_k,
\end{equation}
where $\rho^\mathrm{ac}$ is the shared APs coefficient for the transmit power coefficient of the APs, $f_{mk}^\mathrm{ac}$ and $p_{mk}$ denote the beamforming and power control coefficients of the $m$-th AP for the $k$-th UE, and $q^{\mathrm{ac}}_k$ represents the intended message for the $k$-th UE which satisfies $\bbE\left[\absq{q^{\mathrm{ac}}_k}\right] = 1$. After all the APs prepare their messages, they transmit them to their users. Note that all the APs are assumed to maintain sufficient clock synchronization at the sub-6GHz access channel, which is further facilitated by the adoption of mmWave-based synchronization, as briefly highlighted in \sref{sec:architecture}. Now, if $g_{mk}$ denotes the  access channel coefficient between the $m$-th AP and $k$-th UE, the received signal at the UE $k$ can be expressed as
\begin{equation} \label{eqn:accesschannelsignalmodel}
y^{\mathrm{ac}}_k= \sum_{m=1}^{M} x_m g_{mk} + w_k,
\end{equation}
where $w_k \sim \mathcal{CN}(0, 1)$ is the receive  noise at user $k$.

\textbf{Channel Models:} For the fronthaul channel, we adopt a geometric channel model \cite{heath2016overview, Alrabeiah2019c}. In this model, there are $V$ propagation paths between the CPU and each AP. We denote the large- and small-scale fading and azimuth angle of departure of the path $v$ by $\beta_{m,v}^{\mathrm{fh}}$, $\alpha_{m,v}^{\mathrm{fh}}$ and  $\theta_{m,v}$, respectively. With this notation, we can write the channel as
\begin{equation}
	\bh_m = \sqrt{N} \sum_{v=1}^V \sqrt{\beta^{\mathrm{fh}}_{m,v}} \alpha^{\mathrm{fh}}_{m,v} \ba(\theta_{m,v}),
\end{equation}
where $\ba(\cdot) \in \mathbb{C}^N$ is the array response vector function. For the access channel, we define the channel coefficient between AP $m$ and UE $k$ as $g_{mk}=\sqrt{\beta^{\mathrm{ac}}_{mk}} \alpha^{\mathrm{ac}}_{mk}$, where $\beta^{\mathrm{ac}}_{mk}$  and $\alpha^{\mathrm{ac}}_{mk}$  represent the large- and small-scale fading coefficients.

\section{Problem Formulation: End-to-End Rate Optimization}\label{sec:problemformulation}

In this section, we formulate the optimization problem of the end-to-end achievable rate of the proposed wireless fronthaul based cell-free massive MIMO system. We first present the achievable rates of the access and fronthaul channels. Then, we present a transmission schedule, and formulate the end-to-end rate optimization problem with this schedule. In the following, we omit the superscripts denoting fronthaul and access channel variables as they are easily distinguished from the context.

\subsection{Achievable Rate of the Fronthaul}
Considering the system model in \sref{sec:systemmodel}, the achievable rate of AP $m$, that is part of the $k$th user group $\mathcal{G}_k$, can be written as 
\begin{equation} \label{eqn:fronthaulaprate}
R^\mathrm{fh}_{m}(\bff_k) = \log\bigg(1+{\rho^\mathrm{fh} \absq{ \bh_m^H \bff_k}}\bigg),
\end{equation}
where $\bff_k$ is the beamforming vector used by the CPU to serve the set of APs in  the group $\mathcal{G}_k$. Now, note that the message of the $k$th user is simultaneously transmitted to  all the APs in $\mathcal{G}_k$, and that all the APs in the group should finish receiving the message before they start transmitting it to the user. To incorporate that, we define the effective rate of group $k$ as the minimum rate of the APs in the  $\mathcal{G}_k$'s group. Mathematically, we  write
\begin{equation}
R^{\mathrm{fh}}(\mathcal{G}_k, \bff_k) = \min_{m \in \mathcal{G}_k} \big\{ R^{\mathrm{fh}}_m (\bff_k) \big\}.
\end{equation}
Moreover, due to the TDMA schedule of the AP groups, the group rates will be further scaled by the TDMA time fractions. Let $t_k$ denote the fraction of TDMA time allocated to group $\mathcal{G}_k$. Then, the time-scaled fronthaul rate of group $k$ can be written as $t_k R^{\mathrm{fh}}(\mathcal{G}_k, \bff_k)$.  We note that these time fractions satisfy the constraints $0 < t_k < 1$ and $\sum_{k=1}^K t_k=1$. 

\subsection{Achievable Rate of the Cell-Free Massive MIMO Access Channel}
The access channel adopts the main assumptions of the cell-free massive MIMO architectures in \cite{ngo2017}. Specifically, we assume the following: (i) The APs synchronously serve the UEs without any cell boundaries, i.e., each AP can serve any UE. (ii) Time-division duplexing (TDD) is adopted for the transmissions, which facilitates the estimation of the downlink access channel coefficients through the uplink pilots. (iii) Only the large-scale fading coefficients are available at the CPU for joint power allocation. Further, we assume that  the APs adopt conjugate beamforming for the downlink transmission to the users. More specifically, the uplink pilots are used to estimate the uplink channels (which are also used to construct the downlink channels leveraging channel reciprocity). Then, the information about the large-scale fading coefficients are frequently transmitted to the CPU. The CPU uses this large-scale fading information to determine the access channel power coefficients. The APs adopt this power allocation while jointly serving their users. Next, we first describe the adopted channel estimation procedure in detail before formulating the achievable rates of the access link. 

\textbf{Channel Estimation:} In the adopted protocol, the UEs transmit orthogonal  pilot sequences of length $L_p$, $\bpsi_1, \ldots, \bpsi_K \in \mathbb{C}^{L_p}$,  simultaneously to be received by all the APs. If $\rho_t$ denote the power level selected for the pilot transmissions, then the received signal at AP $m$ can be written as 
\begin{equation}
\by^t_m = \sqrt{\rho_t L_p} \sum_{k=1}^{K} g_{mk} \bpsi_k + \bn_m,
\end{equation}
where $\bn_m \sim \mathcal{CN}(0, \bI_{L_p})$ is the receive noise. With this received signal, the MMSE estimator for $g_{mk}$ can be given by
\begin{equation}
\hat{g}_{mk} = \frac{\sqrt{\rho_t L_p} \beta_{mk}}{1+\rho_t L_p \beta_{mk}} \bpsi_k^H \by^t_m.
\end{equation}
We define the channel estimation error $\tilde{g}_{mk} = g_{mk} - \hat{g}_{mk}$, and note that the estimation $\hat{g}_{mk}$ and the error $\tilde{g}_{mk}$ are uncorrelated thanks to the estimator. Therefore, the distributions of the estimated channel coefficients and the error can be written as $\hat{g}_{mk} \sim \mathcal{CN}(0, \hat{\beta}_{mk}), \quad \tilde{g}_{mk} \sim \mathcal{CN}(0, \beta_{mk}-\hat{\beta}_{mk})$ with the variance of the estimator is defined as
\begin{equation}
\hat{\beta}_{mk}=\frac{\rho_t L_p \beta_{mk}^2}{1+\rho_t L_p \beta_{mk}}.
\end{equation}

\textbf{Achievable Rate:} With the adopted conjugate beamforming, the coefficients $f^\mathrm{ac}_{mn}$ in the received signal equations   \eqref{eqn:accesschannelsignalmodel-a}-\eqref{eqn:accesschannelsignalmodel} can be replaced by $\hat{g}_{mn}^*$. Further, with the described changes on the signal model, the capacity lower bound for the UEs given in \cite{nayebi2017precoding} becomes valid, and we can express the achievable rate of user $k$ as 
\begin{equation} \label{eqn:accesslinkrate}
R^{\mathrm{ac}}_k = \log_2\bigg(1+\text{SINR}_k\bigg),
\end{equation}
\begin{equation}
	\text{SINR}_k = \frac{\rho^\mathrm{ac} \big(\sum_{m=1}^M \sqrt{p_{mk}}\hat{\beta}_{mk} \big)^2}{\rho^\mathrm{ac} \sum_{m=1}^M \beta_{mk} \sum_{k'=1}^K p_{mk'} \hat{\beta}_{mk'} + 1},
\end{equation}
where we note here that the power coefficient $p_{mk}$ for the AP $m$ and UE $k$ is set to $0$ if the AP is not in the group of that UE, i.e., if $m \notin \mathcal{G}_k$.  These power coefficients for each AP $m$ also satisfy 
\begin{equation} \label{eqn:powerconstraint}
\sum_{k=1}^K p_{mk} \hat{\beta}_{mk}\leq 1, \quad  \forall m\in \mathcal{M},
\end{equation} 
which captures the total power constraint of AP $m$. 

\subsection{Transmission Schedule} \label{sec:schedule}
\begin{figure}[!t]
	\centering
	\includegraphics[width=.9\columnwidth]{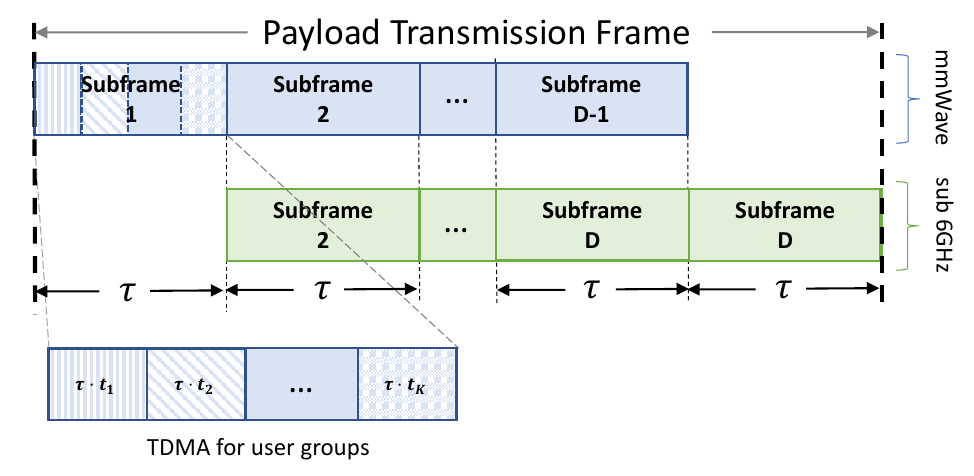}
	\caption{The timing of the fronthaul and access channel transmissions with the frame structure.}
	\label{fig:schedule}
\end{figure}

To achieve the high data rates with proposed transmission scheme, i.e., cell-free massive MIMO access channel and TDMA based fronthaul transmission scheme, a fronthaul transmission schedule needs to be designed. On one hand, the access channel requires the message of each UE to be available at all of the APs in its groups. On the other hand, the fronthaul transmissions are proposed in a TDMA manner, and a set of full transmissions needs to be completed before the access channel transmissions. To that end, we propose a frame/subframe structure that allows almost concurrent data rates. Specifically, we consider a frame of downlink data transmission period within the coherence time, that is split into $D$ subframes. Let the duration of this subframe be $\tau$. Then, in the first subframe of a frame, only the fronthaul transmission is carried out. Then, in the following subframes, the APs transmit the data received in the previous subframe, while receiving the data for the access channel transmissions in the next subframe. With large number of subframes and smaller subframe durations, almost concurrent transmission data rates can be achieved. We illustrate the adopted transmission schedule in \figref{fig:schedule}. Further, with this model, the additional transmission delay of the system can be determined as $\tau$, as the access channel transmissions follow the fronthaul transmissions from the previous subframe. Although this may introduce some delay, the small values can be achieved with a careful fronthaul communication design. Next, we formulate the end-to-end data rate problem with of the proposed system.

\subsection{End-to-End Achievable Rate Optimization}
Based on the presented the transmission schedule and achievable rates for the fronthaul and access channels, we now derive the end-to-end achievable rate of the system. With the adopted system model in \sref{sec:systemmodel} and the schedule given in \sref{sec:schedule}, the access channel rate of UE $k$ during the payload transmission frame can be written as $\frac{D-1}{D} B^{\mathrm{ac}} R^{\mathrm{ac}}_k$, and the fronthaul rate of UE $k$ can be written as $\frac{D-1}{D} t_k B^{\mathrm{fh}} R^{\mathrm{fh}}_k$. As the AP operation is essentially a relaying operation between the fronthaul and access channels, the minimum of these rates provides the end-to-end achievable rate\footnote{This objective is commonly adopted in the relay data rate optimization (e.g., \cite{ikhlef2012max}).}. By dropping $\frac{D-1}{D}$ as $D \rightarrow \infty$, we express end-to-end  rate of UE $k$ by
\begin{equation}\label{eqn:endtoendrate}
	R_k = \min\left\{B^{\mathrm{ac}} R^{\mathrm{ac}}_k,\  t_k B^{\mathrm{fh}} R^{\mathrm{fh}}_k \left(\bff_k\right)\right\},
\end{equation}
Note that the end-to-end channel rate accounts for the fronthaul and access bandwidths, which could be significant since the fronthaul bandwidth in our mmWave-based fronthaul is expected to be much larger than the  bandwidth of the sub-6GHz access channel. Now,  to optimize the end-to-end rate, we adopt the following formulation of the joint max-min fair rate optimization problem 
\begin{subequations} \label{eqn:maxminfairness}
\begin{align}
\max_{\boldsymbol{\mathcal{G}}, \{\bff_k\}, \{t_k\}, \{p_{mk}\}} \,  \min_k \ & \ R_k \\
\quad \mathrm{s.t.}\quad & \sum_{k \in \mathcal{K}} p_{mk} \hat{\beta}_{mk} \leq 1,\quad \forall m \in \mathcal{M} \label{const:1a}\\
\quad &  p_{mk} = 0, \quad \forall m \notin \mathcal{G}_k, \  \forall k \in \mathcal{K} \label{const:1b} \\
\quad & 0<t_k<1,\quad \forall k \in \mathcal{K} \label{const:1c}\\
\quad & \sum_{k \in \mathcal{K}} t_k = 1 \label{const:1d} \\
\quad & \bff_k \in \boldsymbol{\mathcal{F}}, \quad \forall k \in \mathcal{K} \label{const:1e}
\end{align}
\end{subequations}
which aims to jointly optimize the AP groups, the fronthaul beamforming vectors, the fronthaul time allocation for different groups, and the access channel power coefficients. The problem is non-convex and challenging, especially due to the AP grouping and the fronthaul analog-only beamforming. It is worth mentioning here that the AP grouping and analog beamforming can be optimally designed via an exhaustive over all the possible groups and candidate beam codewords, but this will require prohibitive complexity. To reduce this complexity, we propose an iterative sub-optimal solution in the following section. 

\section{Proposed Solution} \label{sec:solution}

In this section, we develop a suboptimal yet efficient solution for the end-to-end data rate maximization problem of the proposed cell-free massive MIMO architecture. We will then show in \sref{sec:numresults} that this developed solution (for the wireless mmWave-fronthauled cell-free massive MIMO architecture) achieves close performance to the upper bound which is given by the fiber-fronthaul based cell-free massive MIMO architecture. To start, we note that the optimization problem in \eqref{eqn:maxminfairness} can be written in the equivalent form 
\begin{align}\label{eqn:maxminfairness2}
\begin{split}
\max_{\boldsymbol{\mathcal{G}}, \{\bff_k\}, \{p_{mk}\}, \{t_k\}} \, \min \ & \left\{B^{\mathrm{ac}} R^{\mathrm{ac}}_k, \, t_k B^{\mathrm{fh}} R^{\mathrm{fh}}_k  (\bff_k) \right\}_{k=1}^K\\
\quad \mathrm{s.t.}\quad & \eqref{const:1a} - \eqref{const:1e}
\end{split}
\end{align}
where the objective is to maximize the minimum of all the fronthaul and access channel rates. 
Before making an attempt for the solution, we highlight the following  remark.
    
    \begin{remark} \label{rem1}
    	For the given system model and the max-min fairness problem defined in \eqref{eqn:maxminfairness2}, the separate optimization of the access channel and fronthaul variables depend on the grouping variable $\boldsymbol{\mathcal{G}}$. Nevertheless, for a fixed grouping $\boldsymbol{\mathcal{G}}$, only the objective function (but not the constraints) retains the variables of both the fronthaul ($\bff_k$ and $t_k$) and access channel ($p_{mk}$), i.e., each constraint affects either the fronthaul channel or the access channel. Therefore, for a given grouping, a two step approach can be developed with the access channel and fronthaul optimization steps to obtain an optimal solution.
    \end{remark}

Motivated by Remark \ref{rem1}, instead of the joint optimization of the fronthaul and access rates, we first consider a given grouping structure, and design two separate problems that maximize the fronthaul and access rates: (i) The access channel power coefficients are optimized without any consideration of the fronthaul, and (ii) the beamforming vectors for each group and TDMA time fractions are determined to maximize the fronthaul rate in a fair manner. As this process is conditioned on the  grouping structure, an optimization of this grouping is required in order to maximize the end-to-end data rates. For that, we propose an iterative heuristic algorithm for the AP grouping where each iteration involves solving two sub-problems for the fronthaul and access rates. The next three subsections present the details of this proposed approach.

\subsection{Access Channel Rate Optimization} \label{subsec:acrate_opt}
As briefly described, for a given AP grouping selection, we first optimize the power coefficients for the access channel, without any fronthaul limitations. This approach allows us to allocate the power of the APs over the access channel, in a similar way to the standard approaches in the cell-free massive MIMO literature. To formulate the access channel optimization problem, we only keep the access channel related terms and constraints of the original problem defined in \eqref{eqn:maxminfairness2}, and write
\begin{subequations} \label{eqn:accesschannel_original}
	\begin{align}
	\max_{\{p_{mk}\}} \,  \min_k \ & \ B^\mathrm{ac} R^\mathrm{ac}_k \\
	\quad \mathrm{s.t.}\quad & \sum_{k \in \mathcal{K}} p_{mk} \hat{\beta}_{mk} \leq 1,\quad \forall m \in \mathcal{M} \\
	\quad &  p_{mk} = 0, \quad \forall m \notin \mathcal{G}_k, \  \forall k \in \mathcal{K},
	\end{align}
\end{subequations}
where two further simplifications can be applied to the objective. First, since $B^\mathrm{ac}$ is a positive constant multiplied with a function of the variables, it can be removed. Second, the objective can be re-written in terms of the SINR values, instead of the rates of the form $\log(1+\textrm{SINR})$. This re-formulation does not change the optimal power coefficients, since $\log(1+\textrm{SINR})$ is a non-decreasing function of the SINR values. Thus, we simplify \eqref{eqn:accesschannel_original} and write the grouping sensitive max-min SINR optimization for the access channel as
\begin{subequations} \label{eqn:accesschannel}
	\begin{align}
	\max_{\{p_{mk}\}} \, \min_k \ & \ \mathrm{SINR}_k \\
	\quad \mathrm{s.t.}\quad & \sum_{k=1}^K p_{mk} \hat{\beta}_{mk} \leq 1,\quad \forall m \in \mathcal{M} \\
	& p_{mk} = 0, \quad \forall m \notin \mathcal{G}_k, \  \forall k \in \mathcal{K} \label{const:grouping},
	\end{align}
\end{subequations}
which is in the same form with the power allocation problem of the standard cell-free massive MIMO \cite{ngo2017} with the addition of the grouping constraint in \eqref{const:grouping}.  In fact,  when all the APs transmit to all UEs, i.e., $\mathcal{G}_k = \mathcal{M}$ $\forall k \in \mathcal{K}$, \eqref{const:grouping} does not provide any constraints, and the presented  optimization of the access channel becomes directly equivalent to the power allocation of cell-free massive MIMO. For the solution of \eqref{eqn:accesschannel}, we first elaborate on the additional grouping constraint \eqref{const:grouping}, which allows an AP to transmit only to the UEs whose groups include that AP, by restricting the power allocated for the other UEs to $0$. It is a linear equality constraint which does not affect the convexity of the problem. Hence, for the problem \eqref{eqn:accesschannel}, the same solution to the optimal power-allocation of the standard cell-free massive MIMO given in \cite{ngo2017} can be applied. Further, different sub-optimal solutions with lower complexity and feedback (e.g., \cite{nayebi2017precoding, interdonato2019ubiquitous, bjornson2020scalable}) can be applied instead of the optimal solution of the problem.

\subsection{Fronthaul Channel Rate Optimization} \label{ssec:fronthaulopt}
In this subsection, we define an optimization problem to maximize the fronthaul rates of the distributed AP groups in a fair manner. First, by starting with the original formulation given in \eqref{eqn:maxminfairness2}, we write a max-min fair fronthaul optimization problem by eliminating the access channel related constraints and objectives, to get 
\begin{subequations} \label{eqn:fronthaulopt}
	\begin{align}
	\max_{\{t_k\} \{\bff_k\}} \min_k &\ t_k R^{\mathrm{fh}}(\mathcal{G}_k, \bff_k)  \\
	\quad \mathrm{s.t.}\quad & \sum_{k \in \mathcal{K}} t_k = 1 \\
	\quad & t_k \geq 0,\quad \forall k \in \mathcal{K} \\
	\quad & \bff_k \in \boldsymbol{\mathcal{F}}, \quad \forall k \in \mathcal{K}.
	\end{align}
\end{subequations}

For this problem in \eqref{eqn:fronthaulopt}, and given the group selections $\left\{\mathcal{G}_k\right\}$, we note that the optimal solution can be obtained by first optimizing the rates of all the groups, $R^{\mathrm{fh}}(\mathcal{G}_k,\bff_k)$, $\forall k$,  and then optimization the TDMA time allocation. This is because any set of TDMA time fractions,  $\left\{t_k\right\}$, does not affect the optimization of the beamforming vectors. Therefore, we will first optimize the beamforming vectors to optimize the group rates individually. Then, we can determine the TDMA time fractions for each group to optimize the time-scaled group rates, as described in the remaining part of this subsection. 

\textbf{Beamforming Optimization} The beamforming vector of each group needs to be optimized to maximize the group rate. Since this rate is determined by the minimum rate of the APs in the group, we can formulate the beamforming optimization problem of any group $\mathcal{G}_k$ as follows 
	\begin{align}\label{eqn:fronthaulgroupopt}
	\begin{split}
	\bff_k^\star= \argmax_{\bff_k} \quad & \left( R^{\mathrm{fh}}(\mathcal{G}_k, \bff_k) = \min_{m \in \mathcal{G}_k} \big\{ R^{\mathrm{fh}}_m (\bff_k) \big\} \right)  \\
	\quad \mathrm{s.t.}\quad & \bff_k \in \boldsymbol{\mathcal{F}},
	\end{split}
	\end{align}
	which coincides with the well-studied  multicast beamforming problem with analog phase-shifters. The optimal solution to the problem can be obtained by exhaustive search which has the complexity increasing exponentially with number of antennas, i.e., $\mathcal{O}(\abs{\mathcal{Q}}^N) = \mathcal{O}(2^{q N})$. To reduce this complexity, several methods have been developed in the literature \cite{wang2019energy, chen2020improved}. In the simulation results in \sref{sec:numresults}, we  adopt the suboptimal iterative solution proposed in \cite{wang2019energy} due to its low complexity and good performance.  With the optimized beamforming vectors, the optimal time fractions of the TDMA can be determined. Next, we will present two approaches to solve this TDMA time fraction optimization problem. 

\textbf{TDMA Optimization. Approach 1:} Given the solution of the problem \eqref{eqn:fronthaulgroupopt},  $\left\{\bff^*_k\right\}$, the fronthaul rate of each group $k$ becomes a constant denoted by $R^{\mathrm{fh}}(\mathcal{G}_k) = R^{\mathrm{fh}}(\mathcal{G}_k, \bff^*_k)$. With these rate constants, we can simplify the fronthaul rate maximization problem given in \eqref{eqn:fronthaulopt} to be
\begin{align} \label{eqn:fronthaulequal}
	\max_{\{t_k\}} \min_k &\ t_k R^{\mathrm{fh}}(\mathcal{G}_k) \ \mathrm{s.t.} \  \sum_{k \in \mathcal{K}} t_k = 1, \ \ t_k \geq 0 \  \forall k \in \mathcal{K}.
\end{align}
As the problem is linear, the optimal solution is obtained at the equality of the time scaled fronthaul rates, i.e., $t_k R^\mathrm{fh}_k = t_k' R^\mathrm{fh}_{k'}$ $\forall k, k' \in \mathcal{K}$. Therefore, the optimal sum of the time scaled group rates can be given by
\begin{equation} \label{eqn:fronthaulrate}
	R^\mathrm{fh}(\boldsymbol{\mathcal{G}}) = \mathrm{HM}\{R^\mathrm{fh}(\mathcal{G}_k)\}.
\end{equation}
where the details of the proof is provided in Appendix A.

Finally, with the obtained access channel and fronthaul rates, we can take the minimum of them to provide an efficient solution to \eqref{eqn:maxminfairness2} for a given grouping structure. 
	
	The provided solution so far only aims to maximize the minimum of the rates, which is also the objective of \eqref{eqn:maxminfairness2}. However, the optimal solution is not necessarily unique, which motivates the search within this set of solutions for one that could further optimize other objectives. In particular, without decreasing the minimum of the end-to-end rates of the given groups, we may still  be able to increase the data rates of the other groups by further optimizing the fronthaul TDMA time allocations of these groups. To detail, the solution given in \eqref{eqn:fronthaulrate} attempts to maximize the minimum of the fronthaul group rates without accounting for the optimized access channel rates. The user rates, however, depend on both the access and fronthaul rates. Therefore, we may be able to further increase the data rates of the users by taking into account the access rates as a restricting constraint.  Based on that, we define the following problem which attempts to optimize the fronthaul rate of each group to meet the access rate of the group in a fair manner.  
	
\textbf{TDMA Optimization. Approach 2:} For a fair allocation, the time-scaled rate of the groups should be equal, unless any of the constraints are met. Therefore, the time allocated to each group should be inversely proportional to their rates. To achieve this, we utilize the weighted logarithm function for the objective, i.e., $\sum_k w_k \log(t_k)$, which allows the resources to be allocated fairly, proportional to the weights, $w_k$. We select the weights as $w_k=\frac{1}{R^{\mathrm{fh}}(\mathcal{G}_k)}$, to allocate the time fractions inversely proportional to the fronthaul rates of the groups, resulting in equal fronthaul rates. In addition, we upper-bound the fronthaul rate of a user by the access channel rate of that user. With this upper-bound, the total time fractions do not necessarily meet the summation equality, $\sum_{k\in \mathcal{K}} t_k = 1$. Hence, we relax the condition by $\sum_{k\in \mathcal{K}} t_k \leq 1$, and write the problem for the time allocation of the TDMA, that maximize the end-to-end rate in a fair manner, as follows
\begin{subequations} \label{eqn:fronthaul_opt}
	\begin{align}
		\max_{\{t_k\}} \quad & \sum_{k \in \mathcal{K}} \frac{1}{R^{\mathrm{fh}}(\mathcal{G}_k)} \log t_k  \\
		\mathrm{s.t.}\quad & \sum_{k \in \mathcal{K}} t_k \leq 1 \\
		\quad & t_k  \leq \frac{B^\mathrm{ac} R^\mathrm{ac}_k}{B^\mathrm{fh} R^{\mathrm{fh}}(\mathcal{G}_k)}, \quad \forall k \in \mathcal{K} \\
		\quad & t_k \geq 0,\quad \forall k \in \mathcal{K}.
	\end{align}
\end{subequations} 
We provide the proof of this problem in Appendix B. The resulting solution takes a similar form with the water-filling solution with upper-bounds. Specifically, the solution can be obtained by finding the maximum individual fronthaul rate (water level), $\eta>0$, that satisfy the following inequality
\begin{equation} \label{eqn:fronthaul_solution1}
	\sum_{k \in \mathcal{K}} t^*_k(\eta)  \leq 1, \ \text{with} \ t_k^*(\eta) = \min\left\{\frac{B^\mathrm{ac} R^\mathrm{ac}_k}{B^\mathrm{fh} R^{\mathrm{fh}}(\mathcal{G}_k)} , \frac{\eta}{R^{\mathrm{fh}}(\mathcal{G}_k)} \right\}.
\end{equation}
For the optimal $\eta$ value, $\eta^*$, we can obtain the optimal time fractions $t_k^*=t_k^*(\eta^*)$. The given solution allocates the fronthaul rates equally among UEs, until the satisfaction of individual access channel rates or the use of the total time. In the special case of all the fronthaul rates being smaller than the access channel rates, it allocates the rates equally, coinciding with \eqref{eqn:fronthaulrate}.

\begin{figure*}[!t]
	\centering
	\includegraphics[width=1.7\columnwidth]{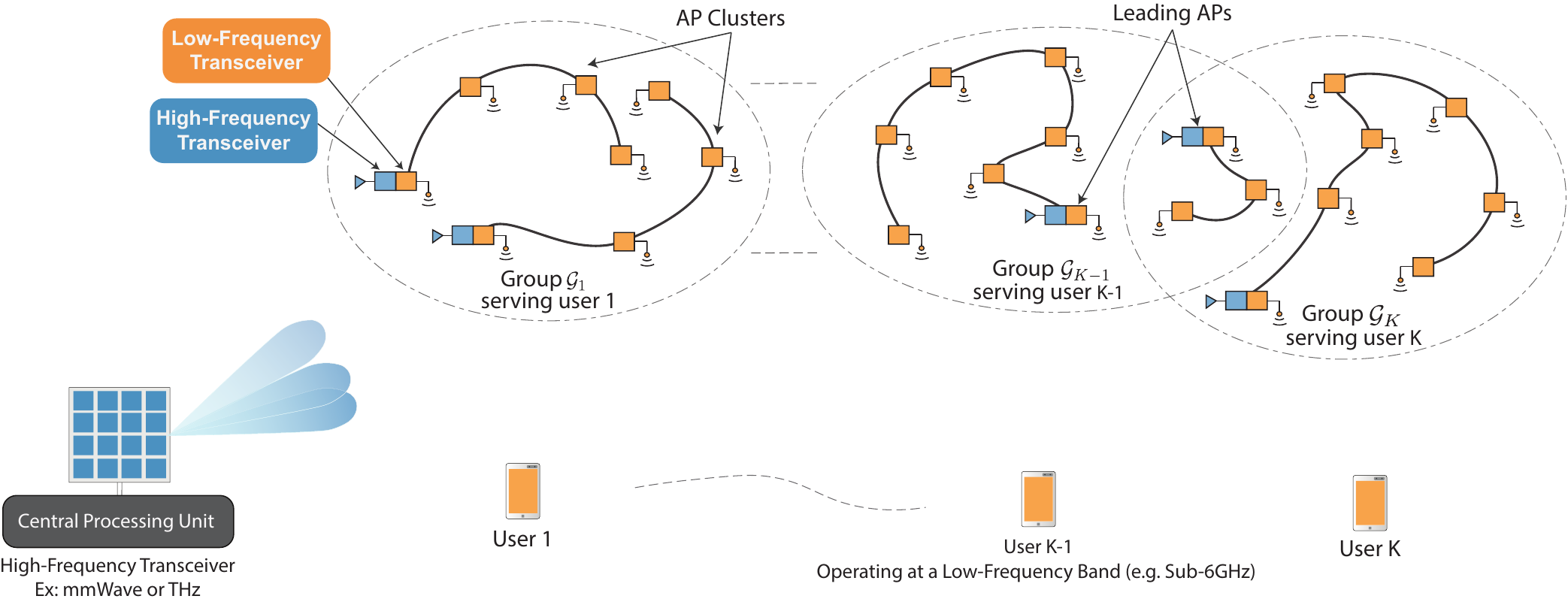}
	\caption{An illustration of the proposed mixed-fronthaul architecture. In this architecture, each subset of APs are connected together via an optical fiber forming a \textit{cluster}. Each cluster has one leading AP that is responsible for the wireless fronthaul communication with the central processing unit. For the access link, each user is served by a group of AP clusters. This architecture has the potential of  reducing the fronthaul cost/complexity while maintaining the data rate/coverage gains as illustrated in \sref{sec:numresults}. }
	\label{fig:systemmodel-connected}
\end{figure*}

\subsection{End-to-End Optimization through Iterative Group Selection} \label{subsec:group}

With the proposed fixed-group end-to-end data rate maximization, the efficient selection of the groups becomes crucial in achieving high data rates. However, the group selection problem is a combinatorial problem that has high complexity given the large number of APs and UEs. In particular, there are $2^M$ possible selection of groups for each UE, leading to a total  of $2^{MK}$ distinct selections for the $K$ users. Moreover, each selection of the groups needs to be utilized with the proposed end-to-end data rate optimization which results in prohibitive optimization complexity for practical systems. Hence, an efficient design for the group selection is required. To this end, we propose a low-complexity yet efficient solution that is motivated by understanding the end-to-end achievable rate optimization problem and the proposed cell-free massive MIMO architecture. Next, we present the proposed approach which has two key points, namely the group selection and the group size optimization.

\textbf{Group Selection:} We adopt the following group selection approach: The group of each user is selected as the $G$ APs with the maximum channel gains. Here, it is important to clarify two points: (i) The group size $G$ is assumed to be fixed for the sake of simplicity and low-complexity solution. Note that the impact of this constraint is expected to be marginal given the high density of the APs. (ii) The selection of the APs with the maximum channel gains will likely lead to a set of APs that are close to each other, which will lead to efficient beamforming design via more focused beams. 
Next, we mathematically describe the proposed approach. Let $\{\beta^{(o)}_{1k}, \ldots, \beta^{(o)}_{Mk}\}$ define  the ordered set of channel coefficients between user $k$ and the $M$ APs. This set adopts a descending order, i.e., the channel coefficients satisfy $\beta^{(o)}_{mk}\geq \beta^{(o)}_{m'k}$ for any $m\leq m'$. In addition, to formalize the mapping between the original channel coefficients sets $\left\{\beta_{1k}, ..., \beta_{Mk}\right\}$ and the ordered channel coefficients set $\{\beta^{(o)}_{1k}, \ldots, \beta^{(o)}_{Mk}\}$, we define the permutation $\varsigma_k(.)$ such that $\beta^{(o)}_{m'k} = \beta_{mk}$ if $\varsigma_k(m') = m$, i.e., if the $m$th AP has the $m'$th highest channel gain. Using these definitions, for a given group size $G$, the group of UE $k$ can be determined as $\mathcal{G}_k = \bigcup_{m'=1}^G \varsigma_k(m')$, which simply selects the $G$ APs with the best channel gains. With this approach, the group selection is reduced to the selection of the parameter $G$, and it needs to be determined carefully.

\textbf{Group Size Optimization:} The number of APs per group, $G$, needs to be optimized to maximize the end-to-end rate. Here, it is interesting to highlight the following trade-off: A small group size may lead to more optimized CPU-APs beamforming design and hence high fronthaul rates. At the same time, it may also result in low APs-user beamforming gain and achievable access channel rates. Therefore, to select $G$, one possible approach is the trial of different group size values from a pre-determined interval (for a given AP structure). Another approach is to start from a certain group size $G$ and then increase/decrease the group size depending on the relation between the access and fronthaul rates. More specifically, if the sum fronthaul rate is larger than the sum access channel rate, i.e., $R^\mathrm{fh} \geq R^\mathrm{acc}$, the group size is increased. Otherwise, if $R^\mathrm{fh} < R^\mathrm{acc}$, the group size is decreased. Through the iterations of these updates, the group value can be locally optimized\footnote{Although the iterative algorithm is applied through the sum data rates, it still achieves some fairness between the users since the group size is fixed.}. 


\section{Mixed-Fronthaul Cell-Free Massive MIMO Architecture} \label{sec:connectedAPs}

In  the previous sections, we proposed and designed a cell-free massive MIMO architecture with wireless higher-band fronthaul.  While this architecture is flexible, scalable,  and  has the potential of reducing the installation cost and time of cell-free massive MIMO systems, it has a few drawbacks: (i) Adding a dual-band relay to each remote wireless AP adds an extra cost to the system and (ii) separately powering the distributed APs may require additional infrastructure cost. To address these points while maintaining the same promising gains, we propose a modified architecture with partially connected APs. In this section, we first describe the proposed architecture in \sref{subsec:mixed_arch_idea} and then briefly present its rate optimization approach in \sref{subsec:mixed_arch_design}. The performance evaluation of this architecture is then detailed in \sref{results_mixed}.

\subsection{Description of the Proposed Mixed-Fronthul Architecture} \label{subsec:mixed_arch_idea}

To reduce the cost of the distributed APs while maintaining the potential deployment/operational gains of using wireless fronthaul, we propose the alternative cell-free massive MIMO architecture depicted in \figref{fig:systemmodel-connected}. In this architecture, every cluster/set of distributed APs is connected with wired connections, e.g., an optical fiber. Further, each set of APs includes one leading AP that has a wireless-fronthaul to the central unit. This leading AP will be responsible for transmitting/receiving the cluster data to/from the central unit. It is worth noting here that the connection between the APs in each cluster could be realized using radio stripes  \cite{interdonato2019ubiquitous}. With this implementation, the proposed mixed-fronthaul cell-free massive MIMO architectures combines the gains of the radio stripes and the installation cost/flexibility gains of the higher-band wireless fronthaul, shaping a promising and practical solution for cell-free massive MIMO systems. 

From an operation perspective, the proposed mixed-fronthaul architecture reduces the number of APs that are simultaneously using the wireless fronthaul, which relaxes the fronthaul requirements in terms of the equipment cost,  the required bandwidth, and the beamforming design. Therefore, this mixed-fronthaul architecture has the potential of further improving the wireless fronthaul rates since better beams can be utilized by the central unit in serving the clusters leading APs. In terms of the end-to-end communication model, the main difference between the  mixed-fronthaul architecture and the original proposed architecture in \sref{sec:architecture} is that all the APs in one cluster communicate with the central unit through one leading AP. Therefore, and to simplify the end-to-end rate optimization, we assume that all the APs in one cluster will serve the same user. This approach is similar to \cite{interdonato2019scalability}, where the user-centric groups for cell-free massive MIMO are selected from the set of APs connected to the same CPU rather than individual APs. If the number of APs in each cluster is large, however, it could be important to relax this constraint, i.e., to allow any user to be served by only a subset of the cluster APs. In the next subsection, we elaborate more on the proposed rate-optimization approach.

\subsection{Proposed Rate Optimization Approach} \label{subsec:mixed_arch_design}

In this subsection, we extend the rate optimization solution presented in \sref{sec:solution} to the modified cell-free massive MIMO architecture with mixed-fronthaul. First, we assume that all the APs in one cluster will be serving the same set of UEs. This is motivated by the negligible data transmission cost within the same cluster.  Hence, the group of APs serving each user $k$ will be determined by selecting the set of clusters (instead of selecting individual APs in the separate-AP architecture in  \sref{sec:architecture}-\sref{sec:solution}). Formally, assume that there are $L$ distinct AP clusters, such that  $\mathcal{C}_l \subseteq \mathcal{M}$, $\forall l\in\{1,\ldots,L\}$ with $\bigcup_{l=1}^L \mathcal{C}_l = \mathcal{M}$ and $\mathcal{C}_l \cap \mathcal{C}_{l'}=\emptyset$, $\forall l \neq l'$. Then, the APs group of user $k$, which we denote as $\mathcal{G}^{\mathrm{cluster}}_k$ to emphasize that they are selected from the available AP  \textit{clusters} will satisfy $\mathcal{G}^{\mathrm{cluster}}_k \subseteq \{1, \ldots, L\}$. Further, we let $a_l$ represent the leading AP of the $l$th cluster. Next, we revisit the fronthaul/access rate optimization and the group selection methodology  from the lens of the  mixed fronthaul architecture.

\textbf{Fronthaul Rate Optimization:} The initial fronthaul rate optimization was defined in \eqref{eqn:fronthaulopt}, which aims to optimize the time and beams of the group $\mathcal{G}_k$. With the mixed architecture, the CPU only needs to transmit the data to the leading APs of the clusters of the $k$-th UE. To simplify the description, we define the user-centric fronthaul groups $\mathcal{G}_k^{\mathrm{fh}} = \{a_l : \ l \in \mathcal{G}^{\mathrm{cluster}}_k\}$ of the leading APs of the clusters. This notation allows a straightforward use of the fronthaul formulation in \eqref{eqn:fronthaulopt}, only by adopting $\mathcal{G}_k^{\mathrm{fh}}$ instead of $\mathcal{G}_k$. With this fronthaul group definition, the rest of the fronthaul optimization methodology can be carried out in the same form.

\textbf{Access Rate Optimization:} Similar to the fronthaul, we let $\mathcal{G}^\mathrm{ac}_k$ denote the set of APs serving each user $k$ over the access channel. This can be  described as  $\mathcal{G}^\mathrm{ac}_k = \{\mathcal{C}_l: \ l \in \mathcal{G}^{\mathrm{cluster}}_k\}$. With this notation, we can utilize the original access channel formulation in \eqref{eqn:accesschannel_original} and the adopted rate optimization approach,  by  replacing the AP groups $\left\{\mathcal{G}_k\right\}$ with   the access channel  groups $\left\{\mathcal{G}^\mathrm{ac}_k\right\}$.

\textbf{Group Selection:} For the selection of the groups, as the user groups are defined in terms of the clusters, a metric for each cluster is needed. For this purpose, we consider the sum of the channel gains of the APs of the clusters, i.e., $\bar{\beta}_{lk} = \sum_{m \in \mathcal{C}_l} \beta_{mk}$, and select the $G$ clusters with the maximum value as the clusters of a UE, $\mathcal{G}^{\mathrm{cluster}}_k$. After determining the active clusters of UEs, the obtained groups of clusters can be converted to the fronthaul and access channel groups of the APs by earlier definitions. Then, the access channel and fronthaul optimization problems can be utilized over the access channel and fronthaul groups, completing an iteration of end-to-end data rate maximization. The described process will evaluated in the numerical results section to show the performance of connected APs.

\section{Numerical Results} \label{sec:numresults}

In this section, we evaluate the performance of the proposed cell-free massive MIMO architectures and the developed rate optimization algorithms. Further, these proposed solutions are compared  with the classical fiber-fronthaul based cell-free massive MIMO systems. The flow of this section is as follows. First, we present the adopted system setup and parameters in \sref{subsec:setup}. Then, we evaluate the performance of the two proposed cell-free massive MIMO architectures, namely with separate APs and connected APs, in Sections \ref{subsec:arch_results} and \ref{results_mixed}.

\begin{table}[!t]
	\centering
	\caption{Simulation Parameters}
	\begin{tabular}{|c|c|c|}
		\hline
		\textbf{Parameter}        & \textbf{Fronthaul} & \textbf{Access Channel} \\ \hline
		Frequency ($f_c^\mathrm{fh}, f_c^\mathrm{ac}$)    & $28$ GHz             & $3.5$ GHz             \\ \hline
		Bandwidth (B)             & $240$ MHz              & $20$ MHz            \\ \hline
		Power ($\tilde{\rho}^\mathrm{fh},\tilde{\rho}^\mathrm{ac}$)      &  $30$ dBm & $10$ dBm      \\ \hline
		Noise Figure ($\sigma^2$) & \multicolumn{2}{c|}{9dB}               \\ \hline
		Antenna Spacing ($d_A$)   & $\lambda/2$        & N/A               \\ \hline
		CPU Antennas ($N$) & \multicolumn{2}{c|}{128}               \\ \hline
	\end{tabular}
	\label{table:simparams}
\end{table}

\subsection{Simulation Setup} \label{subsec:setup}
In the simulation, we considered the proposed cell-free massive MIMO architectures in Section \ref{sec:architecture}  (for the separate APs case) and in \sref{sec:connectedAPs} (for the connected APs case). For both architectures, we adopt mmWave fronthaul channels at $f^\mathrm{fh} = 28$ GHz carrier frequency, and access channels at $f^\mathrm{ac} = 3.5$ GHz. The bandwidth allocated for the access channels is $B^\mathrm{ac} = 20$ MHz, and various values for the fronthaul bandwidth are considered, as will be discussed in the results. The noise figure is set to $9$ dB for both the fronthaul and access links. The central unit is deploying a uniform linear array with $N=128$ elements and half wavelength antenna spacing. For the phase-shift set of analog beamforming at the CPU, we use $q=3$ bit uniform quantizers. The transmit power of the central unit is assumed to be $\tilde{\rho}^\mathrm{fh}=30$ dBm for the fronthaul links while the APs and users transmit powers are set to $\tilde{\rho}^\mathrm{ac}=\tilde{\rho}^t=10$ dBm for the access links and pilot transmissions. The noise power is determined based on the bandwidth of the fronthaul and access channels, following $\sigma_n^2 = \sigma^2 \cdot B \cdot k \cdot T$, with the Boltzmann constant $k$, the channel bandwidth $B$, the temperature $T=\SI{290}{\kelvin}$, and the noise figure $\sigma^2$. The normalized power levels of the different channels are then determined by $\rho = \frac{\tilde{\rho}}{\sigma_n^2}$ with the the corresponding parameters of the specific channel. These baseline system parameters are summarized in Table \ref{table:simparams}. The values that are selected differently from the baseline parameters in different figures are explicitly stated.

At the fronthaul, we consider a LOS multi-path channel with $V=1+V_\textrm{NLOS}$ paths, considering the potential availability of the LOS path with the fixed placement of the CPU and APs. The parameters of the NLOS path are determined randomly, where the number of paths $V_\textrm{NLOS} \sim \mathcal{U}[1, 6]$, angle-of-departure $\theta_{m,v} \sim \mathcal{U}[\pi/2, \pi/2]$ and the small-scale fading $\alpha^\mathrm{fh}_{m, v} \sim \mathcal{CN}(0, 1)$ for $v>1$. The LOS path parameters are determined from the geometry of the simulation model with $\alpha^\mathrm{fh}_{m, 1}=1$.
For the large-scale channel coefficients of LOS and NLOS paths, we adopt the Urban Micro (UMi) street-canyon model given in 3GPP 38.901 \cite{3GPP2017} as follows:
\begin{equation*}
\begin{split}
\mathrm{PL}_\mathrm{LOS}(d, f_c) &= 32.4 + 21\log_{10} d + 21 \log_{10} f_c + \mathcal{X}\\
\mathrm{PL}_\mathrm{NLOS}(d, f_c) &= 32.4  + 31.9\log_{10} d + 21 \log_{10} f_c + \mathcal{X},
\end{split}
\end{equation*}
where $f_c$ is in GHz and $d$ is in meters. $\mathcal{X}$ (dB) is the shadow fading effect determined by a Gaussian random variable with zero mean and standard deviation $4$ and $8.2$ for LOS and NLOS paths of the fronthaul. For the access channel, we adopt $\mathrm{PL}_\mathrm{NLOS}(d_{mk}^\mathrm{ac}, f_c^\mathrm{ac})$, however, apply a spatially correlated shadow fading exactly as described in \cite{ngo2017} with $\delta=0.5$ and decorrelation distance $100$m. Next, we evaluate the performance of the proposed solutions.

\begin{figure*}[t]
	\centering
	\subfigure[Group of 12 APs]{
		\includegraphics[width=.85\columnwidth]{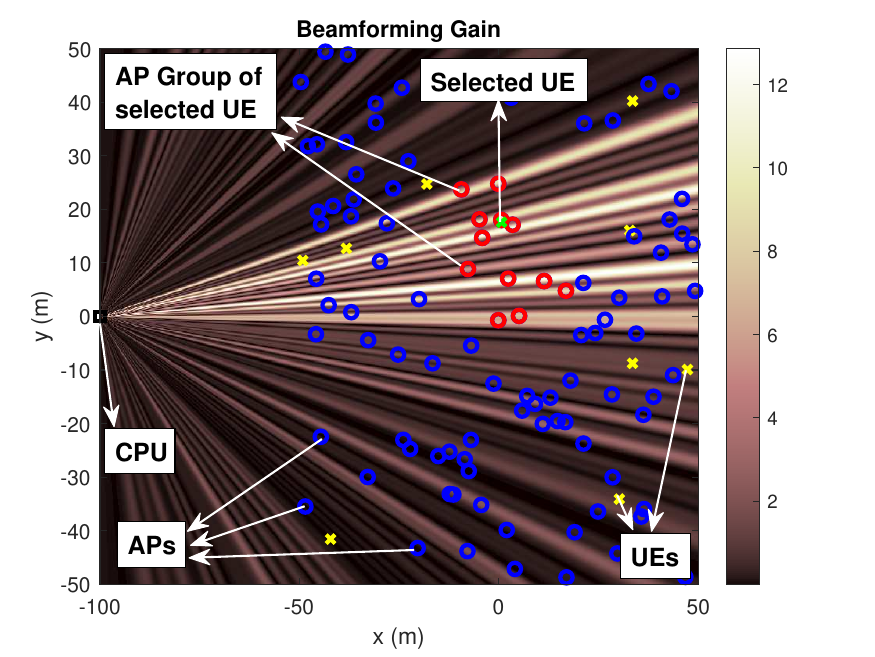}
		\label{fig:optimizedbeams-a}
	}
	\subfigure[Group of 25 APs]{
		\includegraphics[width=.85\columnwidth]{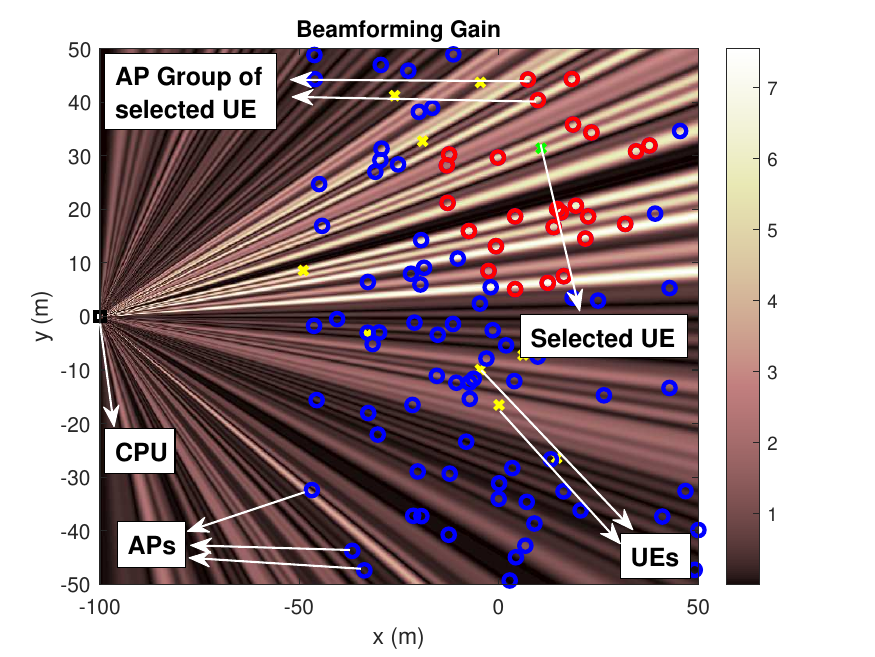}
		\label{fig:optimizedbeams-b}
	}
	\caption{The beamforming gains of the optimized beamforming vectors for the groups of (a) 12 APs and (b) 25 APs. The given scenario adopts randomly placed $M=100$ APs and $K=10$ UEs. The beamforming gain of a group depends on the group size, decreasing with larger groups.} \label{fig:optimizedbeams}
\end{figure*}

\subsection{Evaluating the Proposed Architecture with Separate APs (Wireless-Only Fronthaul)} \label{subsec:arch_results}
The objective of this subsection is to evaluate the performance of the proposed cell-free massive MIMO architecture with wireless fronthaul (and separate APs). In particular, we want to draw some insights into how this architecture performs if compared with the classical fiber-fronthaul based cell-free massive MIMO architecture and whether it can achieve comparable data rates. To do that, we consider a setup where the UEs and APs are randomly placed over an area of $100m \times 100m$ centered at $(0, 0)$. Unless otherwise stated, the following simulations assume that  $K=10$ UEs and $M=100$ APs are placed in the considered area. The central unit (CPU) is located at $\bz_0 = (x_0, y_0) = (-D, 0)$, i.e., at a distance $D=100$m away from the center of the square area. The simulations are averaged over $250$ realizations. Each realization drops the APs and UEs randomly adopting a uniform probability distribution. With this setup, we study the following important questions.

\textbf{What is the performance of the fronthaul beamforming? } To evaluate the performance of the adopted fronthaul beamforming approach\footnote{As mentioned in \sref{ssec:fronthaulopt}, we adopt the suboptimal approach in \cite{wang2019energy} to optimize the beams. The approach starts from a beam vector, and iteratively updates each beamforming coefficient to maximize the rate until the convergence. To ensure a good performance, this solution is applied for $100$ starting points, and the best performing beam is adopted in the simulations.} in our wireless-fronthaul based cell-free massive MIMO architecture, we plot the achievable beamforming gain for the considered deployment in \figref{fig:optimizedbeams}. This figure shows the beamforming gain in x-y coordinates for two different AP group sizes, namely a group of 12 APs in \figref{fig:optimizedbeams-a} and a group of 25 APs in \figref{fig:optimizedbeams-b}. These figures provide a visual verification that the central unit focuses it multi-cast beam towards the APs in the group of interest. In terms of beamforming gain, it varies from 12 (i.e., around 11dB) in the case of 12 APs group to  4 (i.e., around 6dB) in the 25 APs/group case. This highlights the trade-off between the fronthaul rate and access rate as the group sizes increases, as this decreases the fronthaul rate while increasing the access rate.   It is worth mentioning here that this fronthaul beamforming can be further improved when adopting hybrid precoding approaches \cite{Li2019a,Alkhateeb2014c}. Also, leveraging machine learning could enable the autonomous design of these beamforming vectors/codebooks \cite{Alrabeiah2020c,Zhang2021b,Zhang2021a}. This can further  reduce the deployment overhead. 

\begin{figure*}[t]
	\centering
	\subfigure[Sum fronthaul and access channel rates]{
		\includegraphics[width=.85\columnwidth]{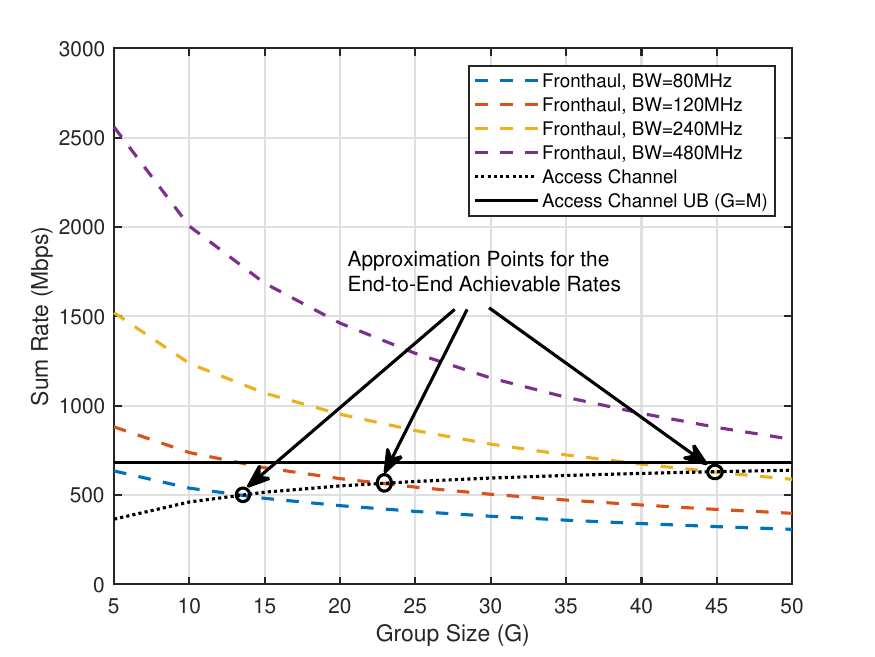}
		\label{fig:rate-a}
	}
	\subfigure[Sum fronthaul, access channel and end-to-end rates]{
		\includegraphics[width=.85\columnwidth]{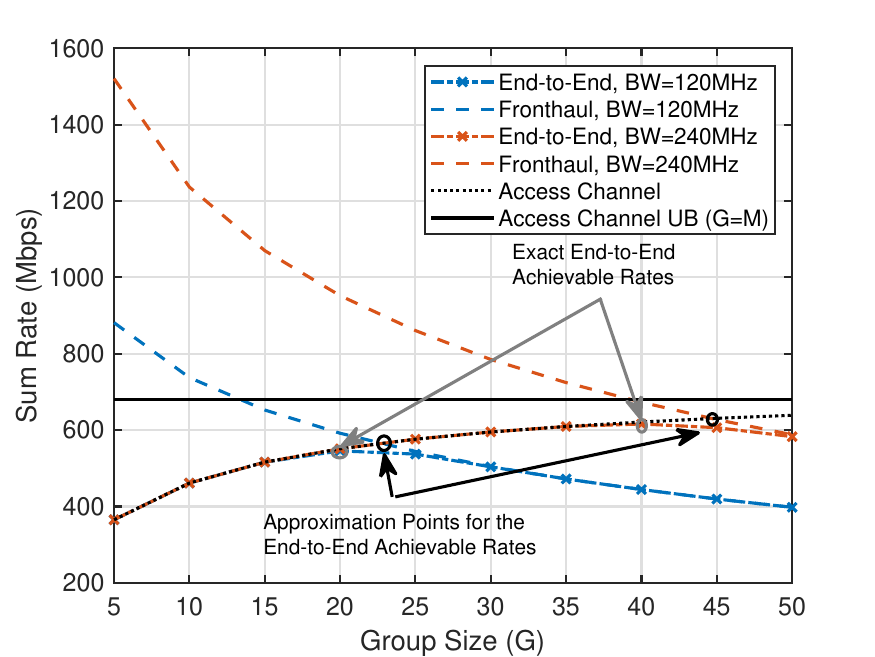}
		\label{fig:acmatching}
	}
	\caption{The access channel and fronthaul sum data rates of the proposed architecture with different group size and fronthaul bandwidth values. (a) shows the change in the fronthaul rates based on the available bandwidth an group size. (b) includes the sum of the end-to-end data rates reflecting the exact performance.
	} \label{fig:rate}
\end{figure*}

\textbf{What are the achievable fronthaul and access rates?} 
Now, we want to evaluate the fronthaul and access achievable rates using the developed rate optimization solutions in \sref{sec:solution}. Adopting the same setup described earlier in this section with $10$ UEs and $100$ APs, we plot the achievable fronthaul/access sum-rates in \figref{fig:rate-a} versus the AP group size. These achievable sum-rates (sum of the rates of UEs) are obtained from the solutions of the fronthaul (Approach 1) and cell-free massive MIMO solutions, and present the upper-bounds on the end-to-end rate. They reflect the behavior of the system within a transmission frame and their intersection point can be considered as a good approximation for the achievable end-to-end sum-rate, as will be discussed shortly in \figref{fig:acmatching}. The group for each user is determined based on the channel gain criterion described in \sref{subsec:group}, the access rate is optimized based on the solution presented in \sref{subsec:acrate_opt}, and the fronthaul time allocation optimization is implemented based on Approach 1 in \ref{ssec:fronthaulopt}. As an upper bound on the cell-free massive MIMO system, we show fully activated APs, i.e., $G=M=100$, without any fronthaul limitations. This curve is abbreviated as Access Channel UB in the figure. As shown in \figref{fig:rate-a}, as the group size increases, the access rate increases because of the higher access channel beamforming gain and the fronthaul rate decreases because of the multicasting to more APs with a lower beamforming gain. With high enough fronthaul bandwidth, e.g., $240$MHz, the proposed rate optimization approach and cell-free massive MIMO architecture intersects at $630$Mbps sum-rate. This is very close to the $680$Mbps sum-rate achieved with $100$ active APs without any fronthaul limitations. These results highlight the promising data rates of the proposed rate optimization solutions.

\textbf{Can the proposed architecture with wireless fronthaul approach the data rates of the classical fiber-fronthaul based cell-free massive MIMO architecture?} 
This is a key question that we are trying to answer in this paper. Essentially, the proposed architecture with wireless higher-band fronthaul has clear gains in terms of  the installation cost/time/flexibility, but could it also provide comparable achievable rates to the fiber-based architecture? In \figref{fig:rate-a}, we provided some insights into the answer of this question by separately evaluating the achievable fronthaul and access rates. The exact achievable rate, however, is the one with matched fronthaul and access rates. To evaluate this exact rate, we adopt the fronthaul rate optimization (Approach 2) in \sref{ssec:fronthaulopt} which ensures that the fronthaul rate of each link does not exceed its corresponding access rate. In \figref{fig:acmatching}, we plot the achievable rates for the access and fronthaul links using Approach 2. This rate then represents the exact end-to-end achievable rates using the proposed architecture. For reference, we also plot the fronthaul rates using Approach 1. As shown in \figref{fig:acmatching}, with sufficient fronthaul bandwidth, the exact achievable end-to-end data rates using the proposed architecture are very comparable to the classical fiber-based cell-free massive MIMO architecture. For example, with $240$MHz fronthaul bandwidth, the proposed solution achieves around $620$Mbps sum-rate compared to $680$Mbps for the fully active APs without any fronthaul limitations. This difference can be further reduced by increasing the fronthaul bandwidth. It is also good to note that the exact rate is very close to the approximation discussed in \figref{fig:rate-a}, i.e., $630$.

\begin{figure*}[t]
	\centering
	\subfigure[Achievable rates with varying number of APs]{
		\includegraphics[width=.85\columnwidth]{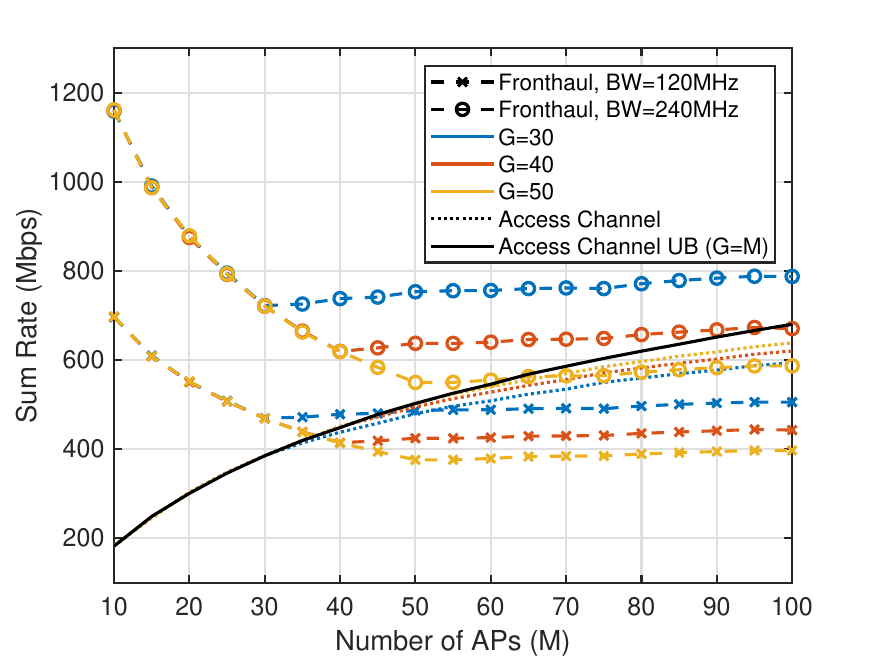}
		\label{fig:scaling-AP}
	}
	\subfigure[Achievable rates with varying number of UEs]{
		\includegraphics[width=.85\columnwidth]{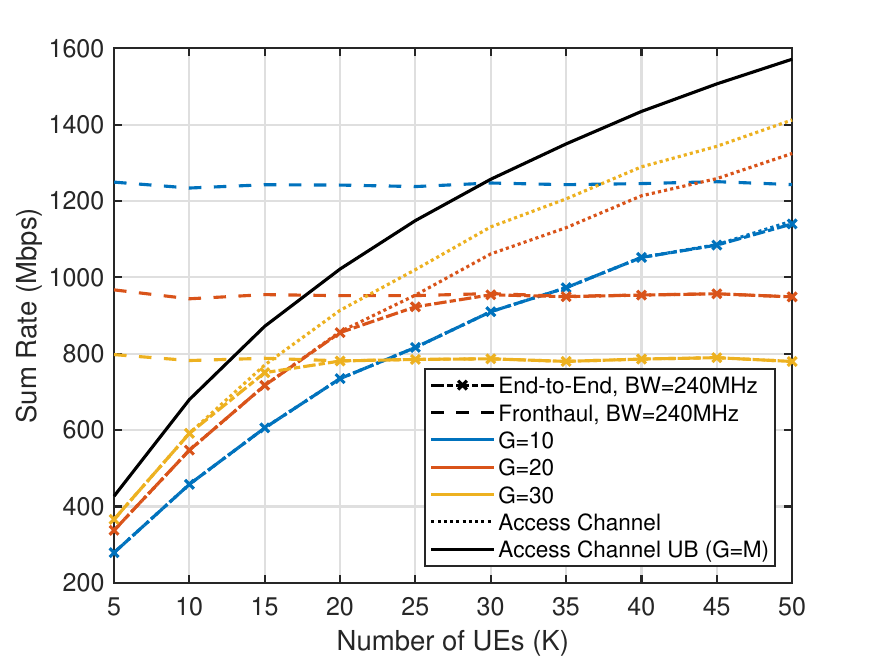}
		\label{fig:scaling-user}
	}
	\caption{The access channel and fronthaul sum data rates of the proposed architecture with different number of (a) APs with $K=10$ UEs, and (b) UEs with $M=100$ APs. The figures show the relationship of group sizes, $G$, to different number of APs and UEs.
	} \label{fig:scaling}
\end{figure*}

\textbf{Can the proposed architecture support a large number of distributed APs and UEs?} The scalability is a key objective of cell-free massive MIMO systems to be able to support more users/larger areas and to increase the beamforming gains/achievable rates.  To draw some insights into the capability of the proposed architecture in supporting large numbers of APs and UEs, we plot the fronthaul/access rates vs  the number of APs in  \figref{fig:scaling-AP}, and the number of UEs in  \figref{fig:scaling-user}.  As shown in \figref{fig:scaling-AP}, for each group size, the access rates increase with denser APs due to the higher access beamforming gains. At the fronthaul, the rate decreases rapidly for $M\leq G$. In this region, $M\leq G$, the group size is taken as $G=M$. Therefore, every new AP introduced into the system also increases the effective group size by adding APs to the groups. This decreases the fronthaul beamforming gains. For larger values of $M$, the fronthaul rates slightly increase, as the area containing a single AP group becomes smaller, allowing higher beamforming gains. Overall, however, the end-to-end data rates, which are approximately given by the  intersections of the fronthaul and access channel rates, are increasing with more APs. This is also very comparable to the increase in data rates experienced by the fiber-based architecture, which highlights the potential of the proposed solution. In \figref{fig:scaling-user}, we investigate the effect of increasing number of UEs. In the figure, the sum fronthaul rate with a fixed group size is almost constant for different number of UEs, mainly due to the fronthaul resources and TDMA based communication design. However, with smaller group size values, the achievable maximum fronthaul rate increases, while less number of APs serve each UE and the access channel rate decreases. At the access channel, with the scaling of UEs, the effect of the group size increases. However, even with $K=50$ APs, the $80\%$ of the data rate with fiber-fronthaul can be obtained while only using $240$MHz fronthaul bandwidth.

\begin{figure*}[t]
	\centering
	\subfigure[Group size of 10 clusters]{
		\includegraphics[width=.85\columnwidth]{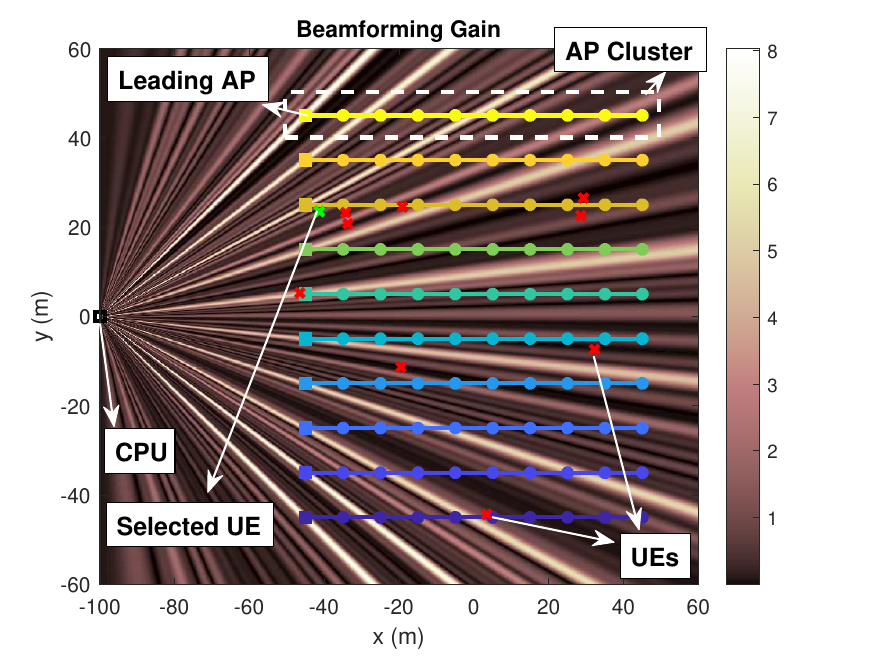}
		\label{fig:beamconnecteda}
	}
	\subfigure[Group size of 9 clusters]{
		\label{fig:beamconnectedb}
		\includegraphics[width=.85\columnwidth]{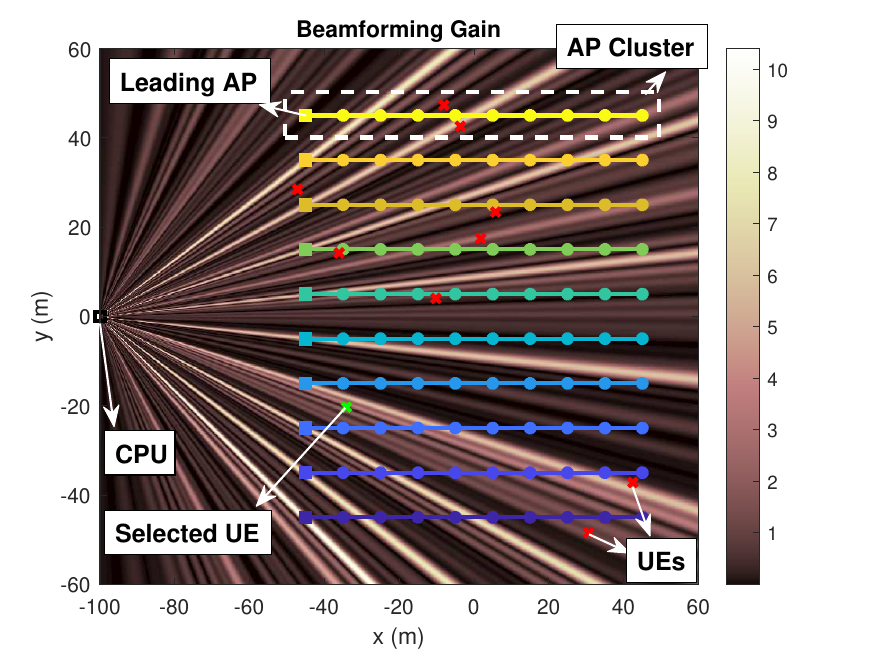}
	}
	\caption{The beamforming gains with the optimized beamforming vectors are shown for the mixed-fronthaul architecture. In (a), all of the clusters are activated, hence the beam gain is maximized towards all of the leading APs. In (b), the bottom $9$ clusters are activated, and the beam gain is not maximized for the leading AP of the top cluster.} 
	\label{fig:beamconnected}
\end{figure*}

\subsection{Evaluating the Proposed Architecture with Connected APs (Mixed-Fronthaul)} \label{results_mixed}

In this subsection, we evaluate the performance of the proposed modified architecture in  \sref{sec:connectedAPs} with connected APs.  As shown in \figref{fig:beamconnected}, we consider a square-grid of APs, where the APs are located uniformly in an $10 \times 10$ grid with equal distance between the rows and columns. For simplicity, we assume that the 10 APs in each line in the $y$-axis form a cluster. The leaders of the clusters (that communicate with the central unit through the mmWave fronthaul) are selected as the closest APs to the central unit. Further, we assume that $K=10$ UEs are randomly located in the $100m \times 100m$ area of interest.  The central unit is located at the origin, as depicted in \figref{fig:beamconnected}, and is equipped with a ULA of $N=128$ antenna elements.

\begin{figure}[t]
	\centering
	\includegraphics[width=.9\columnwidth]{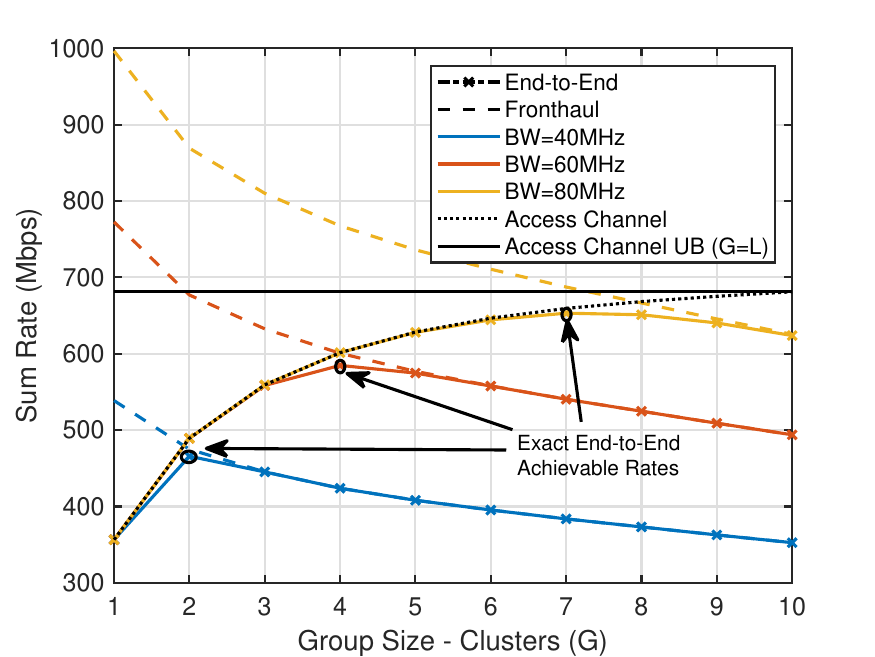}
	\caption{For the mixed-fronthaul architecture, the fronthaul, access channel and end-to-end sum rates are shown for different number of group size (of clusters). In the figure, the $10\times10$ APs rectangular grid scenario illustrated in \figref{fig:beamconnected} is adopted and each cluster consists of $10$ APs. }
	\label{fig:datarate_conn}
\end{figure}

\textbf{When does the achievable data rate with the proposed architecture approach the optimal solution?} 
To evaluate the achievable data rates with the proposed connected-APs cell-free massive MIMO architecture, we adopt the setup described in the previous paragraph, which is also depicted in \figref{fig:beamconnected}. For this setup, we first investigate the achievable beamforming gains for different  numbers of AP clusters per group. To illustrate, we consider one UE and plot the beamforming gains for the clusters serving this UE in two scenarios: (i) When a group of 10 AP clusters (i.e., all the clusters) jointly serve this UE as shown in \figref{fig:beamconnecteda} and (ii) when a group of the 9 bottom clusters serve the UE, as shown in \figref{fig:beamconnectedb}. As expected, when the number of AP clusters per group increases, the beamforming gain decreases. Next, we calculate the achievable fronthaul and access rates of the same setup in \figref{fig:datarate_conn} for different numbers of AP clusters per group. As shown in this figure, the modified architecture with connected APs has clear data rate gains compared to the separate APs architecture studied in \figref{fig:rate}. With only $80$MHz fronthaul bandwidth, the  proposed connected-APs architecture  achieves nearly the same  end-to-end rates obtained by the upper bound, which is defined by the classical fiber-fronthaul based cell-free massive MIMO architecture with fully active APs. This provides a promising solution for flexible, scalable, and efficient cell-free massive MIMO systems.

\section{Conclusions}\label{sec:conclusion}
In this work, we proposed two wireless-fronthaul based architectures for cell-free massive MIMO systems where the fronthaul link is operating at a higher band (e.g., mmWave) compared to the access band (e.g., sub-6GHz). We formulated the end-to-end data rate optimization problems for these architectures based on user-centric AP grouping. Then, we  developed efficient transmission and resource allocation strategies that can achieve near-optimal performance. Simulation results showed that the  proposed architectures can achieve data rates that approach the optimal rates obtained with optical fiber-based fronthaul with reasonable assumptions on the number or CPU antennas, coverage area, number of users, etc. The results also showed that these architecture are capable of supporting large numbers of APs under realistic fronthaul bandwidth requirements. For the future work, it is interesting to develop more advanced fronthaul beamforming solutions using hybrid analog/digital architectures. In addition, more practical studies adopting more realistic cell-free constraints (e.g., pilot contamination), practical issues of the mmWave systems, and energy efficiency of the architecture are interesting directions for future extensions. Further, the design of machine learning based adaptive AP grouping and beamforming optimization approaches for the proposed wireless-fronthaul based cell-free massive MIMO architectures can be investigated. 

\appendices

 \section{} 
For the solution of the problem in \eqref{eqn:fronthaulequal}, we first introduce a slack variable \cite{boyd2004convex}, $\eta$, to represent minimum time-scaled data rate of the groups and write the equivalent formulation
\begin{equation}\label{eqn:harmonicmeanproblem}
\begin{split}
	\max_{\{t_k\}, \eta} \quad  \eta \quad \mathrm{s.t.}\quad  &t_k R^{\mathrm{fh}}(\mathcal{G}_k) \geq \eta \ \forall k \in \mathcal{K},
	 \\&  \sum_{k \in \mathcal{K}} t_k = 1, \ \ t_k \geq 0 \ \ \forall k \in \mathcal{K}.
\end{split}
\end{equation}
This formulation is a linear program and the solution can be attained by finding maximum $\eta$. Let us assume that there exists an optimal $\eta$, $\eta^\star$, that satisfies the minimum rate constraint of \eqref{eqn:harmonicmeanproblem}, i.e.,  $t_k R^{\mathrm{fh}}(\mathcal{G}_k) \geq \eta$. Due to the maximization objective, this constraint is satisfied at the equality, and it is possible to obtain the exact solution thanks to the linear structure of the problem. Hence, for the equality, we can write $\eta^\star = t^\star_k R^\mathrm{fh}(\mathcal{G}_k)$, where $t_k^\star$ denotes the optimal time fraction for group $k$. Then, we have $t^\star_k = \frac{\eta^\star}{R^\mathrm{fh}(\mathcal{G}_k)}$. By replacing $t_k$ with $t^\star_k$ in the time constraint of \eqref{eqn:harmonicmeanproblem}, $\sum_{k \in \mathcal{K}} t_k = 1$, we obtain
 \begin{equation}
 	\eta^\star = \frac{1}{\sum_{k\in\mathcal{K}} \frac{1}{R^\mathrm{fh}(\mathcal{G}_k)}} = \frac{\mathrm{HM}\{R^\mathrm{fh}(\boldsymbol{\mathcal{G}})\}}{K}
 \end{equation}
 which leads to the solution given in \eqref{eqn:fronthaulrate}.

 \section{} 
The problem given in \eqref{eqn:fronthaul_opt} is in the form
	\begin{align}
	\begin{split}
		\max_{\{t_k\}} \quad &\sum_{k \in \mathcal{K}} w_k \log t_k  
		\\ \mathrm{s.t.}\quad &\sum_{k \in \mathcal{K}} t_k \leq 1, \ t_k  \leq \bar{t}_k \ \forall k \in \mathcal{K}, \ \ t_k \geq 0 \ \forall k \in \mathcal{K}
	\end{split}
	\end{align}
for the constants defined by $w_k=\frac{1}{R^{\mathrm{fh}}(\mathcal{G}_k)}$ and $\bar{t}_k =  \frac{B^\mathrm{ac} R^\mathrm{ac}_k}{B^\mathrm{fh} R^{\mathrm{fh}}(\mathcal{G}_k)}$. The problem is concave due to the sum of logarithms in the objective, and it can be solved by the Karush-Kuhn-Tucker (KKT) conditions \cite{boyd2004convex}. For the solution, we first write the Lagrangian function\footnote{We ignore the constraint $t_k \geq 0$ since the logarithm is only defined in this region, and any solution obtained from the formulation will satisfy the condition.}
as
\begin{equation}
	\mathcal{L}(\bt, \boldsymbol{\lambda}, \nu) = \sum_{k \in \mathcal{K}} w_k \log t_k - \nu (\sum_{k\in\mathcal{K} }  t_k - 1)  - \sum_{k \in \mathcal{K}} \lambda_k (t_k - \bar{t}_k),
\end{equation}
where $\boldsymbol{\lambda} = [\lambda_1, \ldots, \lambda_K]$ and $\nu$ are the Lagrange multipliers. We note that the partial derivative of the Lagrangian function is $\frac{\partial\mathcal{L}(\bt, \boldsymbol{\lambda}, \nu)}{\partial t_k} =  \frac{w_k}{t_k} - \nu - \lambda_k$, and write the KKT conditions as
\begin{align*}
	&\frac{w_k}{t_k^\star} - \nu - \lambda_k = 0, \tag*{(Stationarity)} \\
	&\sum_{k \in \mathcal{K}} t_k^\star \leq 1, \quad t_k^\star  \leq \bar{t}_k, \quad \lambda_k, \nu \geq 0, \tag*{(Feasibility)} \\
	&\lambda_k (t_k^\star - \bar{t}_k) = 0, \quad \nu (\sum_{k \in \mathcal{K}} t_k^\star - 1) = 0.  \tag*{(C. Slackness)} \\
\end{align*}
From the stationarity condition, we have $\nu = \frac{w_k}{t_k} - \lambda_k$. Combining this with the complementary slackness condition of $\lambda_k$, we obtain
\begin{equation}
	t^\star_k = 
	\begin{cases}
		\frac{w_k}{\nu} & \text{if } \lambda = 0, \\
		\bar{t}_k & \text{if } \lambda>0.
	\end{cases}
\end{equation}
which can be simplified to $t^\star_k = \min\{\frac{w_k}{\nu}, \bar{t}_k\}$. Finally, we define $\eta=\frac{1}{\nu}$, and place $t_k^\star$ in terms of $\eta$ in the first (primal) feasibility condition, and obtain the solution given in \eqref{eqn:fronthaul_solution1}.
 
\balance

\end{document}